\documentclass[preprint,3p,10pt]{elsarticle}
\usepackage[utf8]{inputenc}
\usepackage[T1]{fontenc}
\usepackage[english,francais]{babel}

\journal{CR Physique}

\usepackage{graphicx}

\usepackage{color}
\usepackage[x11names,svgnames]{xcolor}

\usepackage[T1]{fontenc}
\usepackage{csquotes}
\usepackage[kerning=true,babel=true]{microtype} 

\usepackage{amsmath}
\usepackage{amssymb}
\usepackage{stmaryrd}

\usepackage[english,francais]{babel}

\usepackage[unicode]{hyperref}
\hypersetup{
    colorlinks=true,
}
\def\og{\leavevmode\raise.3ex\hbox{$\scriptscriptstyle\langle\!\langle$~}}
\def\fg{\leavevmode\raise.3ex\hbox{~$\!\scriptscriptstyle\,\rangle\!\rangle$}}

\begin{document}
\hypersetup{
    citecolor=PaleGreen4!80!black,
    linkcolor=DarkRed, 
    urlcolor=DarkSeaGreen4!90!black}
\centerline{Quantum simulation / Simulation quantique}
\begin{frontmatter}


\selectlanguage{english}
\title{Many-body localization: an introduction and selected topics}


\selectlanguage{english}
\author{Fabien Alet}
\ead{alet@irsamc.ups-tlse.fr}
\author{Nicolas Laflorencie}
\ead{laflo@irsamc.ups-tlse.fr}

\address{Laboratoire de Physique Th\'eorique, Universit\'e de Toulouse, CNRS, UPS, France}
\begin{abstract}
{
What happens in an isolated quantum system when both disorder and interactions are present? Over the recent years, the picture of a non-thermalizing phase of matter, the many-localized phase, has emerged as a stable solution. We present a basic introduction to the topic of many-body localization, using the simple example of a quantum spin chain which allows us to illustrate several of the properties of this phase. We then briefly review the current experimental research efforts probing this physics. The largest part of this review is a selection of more specialized questions, some of which are currently under active investigation. We conclude by summarizing the connections between many-body localization and quantum simulations.
}
\vskip 0.5\baselineskip

\selectlanguage{francais}
\noindent{\bf R\'esum\'e}
\vskip 0.5\baselineskip
\noindent
Que se passe-t-il dans un système quantique isolé lorsque celui-ci présente du désordre et des interactions entre particules? Au cours des dernières années a émergé l'image comme solution stable d'une phase qui ne thermalise pas, la phase localisée à N corps (notre traduction de {\it many-body localization}). Nous présentons une introduction simple à la localisation à N corps, \`a travers l'exemple d'une chaîne de spins quantiques, ce qui permet d'illustrer plusieurs propriétés de cette phase. Nous effectuons ensuite une brève revue des efforts expérimentaux actuels cherchant \`a sonder cette physique. La plus grande partie de cette revue est consacrée à une sélection de questions plus spécialisées, la plupart actuellement en cours d'études. Nous concluons en résumant les liens entre localisation à N corps et simulations quantiques.
} \vskip 0.5\baselineskip
\noindent{\small{\it Mots-cl\'es~:} Localisation à N corps, Thermalisation, Simulations, Intrication }
\end{abstract}
\end{frontmatter}
\tableofcontents
\vskip 1cm 
\selectlanguage{english}
In this review article, we aim at presenting an introduction to the rich and strongly active field of many-body localization (MBL) in isolated, interacting quantum systems in presence of disorder. This topic has been tremendously explored over the past few years, covering different areas of physics: quantum chaos, condensed-matter, quantum information, out-of-equilibrium statistical mechanics, ultracold atoms, etc. The global aim is to provide answers to the fundamental question regarding the fate of isolated quantum systems when both disorder and interactions are present. Understanding the interplay between these two key ingredients addresses several fundamental questions on how a quantum system thermalizes (or fails to), hence touching the very foundations of statistical mechanics.

Providing an exhaustive overview of the flourishing literature on MBL would be an exacting task. There are excellent reviews which give a general introduction to the topic~\cite{nandkishore_many-body_2015,altman_universal_2015} as well as a more recent one~\cite{abanin_recent_2017} which furthermore introduces novel developments. Here we shall rather first give a bird’s eye view introduction to MBL, through a specific example, providing references to more specialized reviews and articles when necessary. We then make focuses on several specific points of interest which are currently active research areas, trying to give a general perspective. We also try to keep the discussion simple, meaning that sometimes we do not achieve the level of rigor or completeness, preferring to refer to the relevant literature. We clearly do not pretend to give an exhaustive overview and some aspects of the MBL problem will not be covered.

The more precise plan of this paper is the following. In Sec.~\ref{sec:intro}, we introduce the topic of MBL through the lens of a spin chain example. In particular, we present several hallmarks observables which allow us to distinguish the two different behaviors of many-body localization and thermalization. We then discuss two important aspects of MBL physics, namely the existence of local integrals of motion in the MBL phase (Sec.~\ref{sec:liom}), as well as the possibility of a many-body mobility edge (Sec.~\ref{sec:edge}). Most of the work on MBL has focused on theoretical studies of one-dimensional lattice models, where a stable MBL phase can be shown to exist. We finish this introduction by providing a non-exhaustive list of lattice models where MBL has been sought for/found (Sec~\ref{sec:models}). 

Sec.~\ref{sec:exp} is dedicated to an overview of the experiments that probe MBL physics, mostly in cold-atomic/trapped ions systems.

The long Sec.~\ref{sec:focus} focuses in more details on several current research directions on the topic:  we first discuss the role of symmetries (which symmetries are compatible with/can be broken in MBL states?) in Sec.~\ref{sec:sym} and the key issues of the nature transition to the MBL phase~(Sec.~\ref{sec:transition}) and of entanglement (Sec.~\ref{sec:ent}) in MBL physics. We then address the actual role of disorder (which type? Is disorder needed?) in MBL physics (Sec.~\ref{sec:disorder}), and the partly related issue of anomalous slow transport in one-dimensional systems {\it before} the MBL transition (Sec.~\ref{sec:sub}). We also briefly discuss expectations related to the presence of MBL in dimension larger than one in Sec.~\ref{sec:dim}. Sec.~\ref{sec:randomgraphs} and Sec.~\ref{sec:distinguish} try to provide connections with Anderson localization. In Sec.~\ref{sec:challenge}, we put emphasis on solutions which have been proposed to solve the computational challenges brought by the MBL problem. In this long list of focuses, we also try to include the open questions in the field. Finally, as this is the main subject of this volume, we present the connections between many-body localization and quantum computations in Sec.~\ref{sec:qc}, before concluding in Sec.~\ref{sec:conc}.  

\section{Introduction to MBL physics}
\label{sec:intro}

\subsection{A spin chain example}
As there are already several excellent general introductions to the physics of MBL~\cite{nandkishore_many-body_2015,altman_universal_2015,abanin_recent_2017}, we rather start directly by an example, from which we will illustrate several properties of the two main paradigms for the long-time behavior of disordered quantum systems. Consider the following one-dimensional lattice model, the spin $1/2$ Heisenberg chain in a random magnetic field
\begin{equation}
\label{eq:rfH}
H_{\rm rfH} = \sum_{i} (S^x_i S^x_{i+1} + S^y_i S^y_{i+1} + S^z_i S^z_{i+1}) - \sum_i h_i S^z_i
\end{equation}
with $S^\alpha=\sigma^{\alpha}/2$ where $\sigma^{\alpha=x,y,z}$ are Pauli matrices, and the fields $h_i$ are drawn from a random distribution, with the strength of the disorder denoted by $h$ (as an illustration, take the box distribution $h_i \in [-h,h]$). This model has a $U(1)$ symmetry as it conserves the total magnetization $S^z=\sum_i S_i^z$ even in the presence of disorder. 

\begin{figure}
\begin{center}
\includegraphics[width=0.85\hsize]{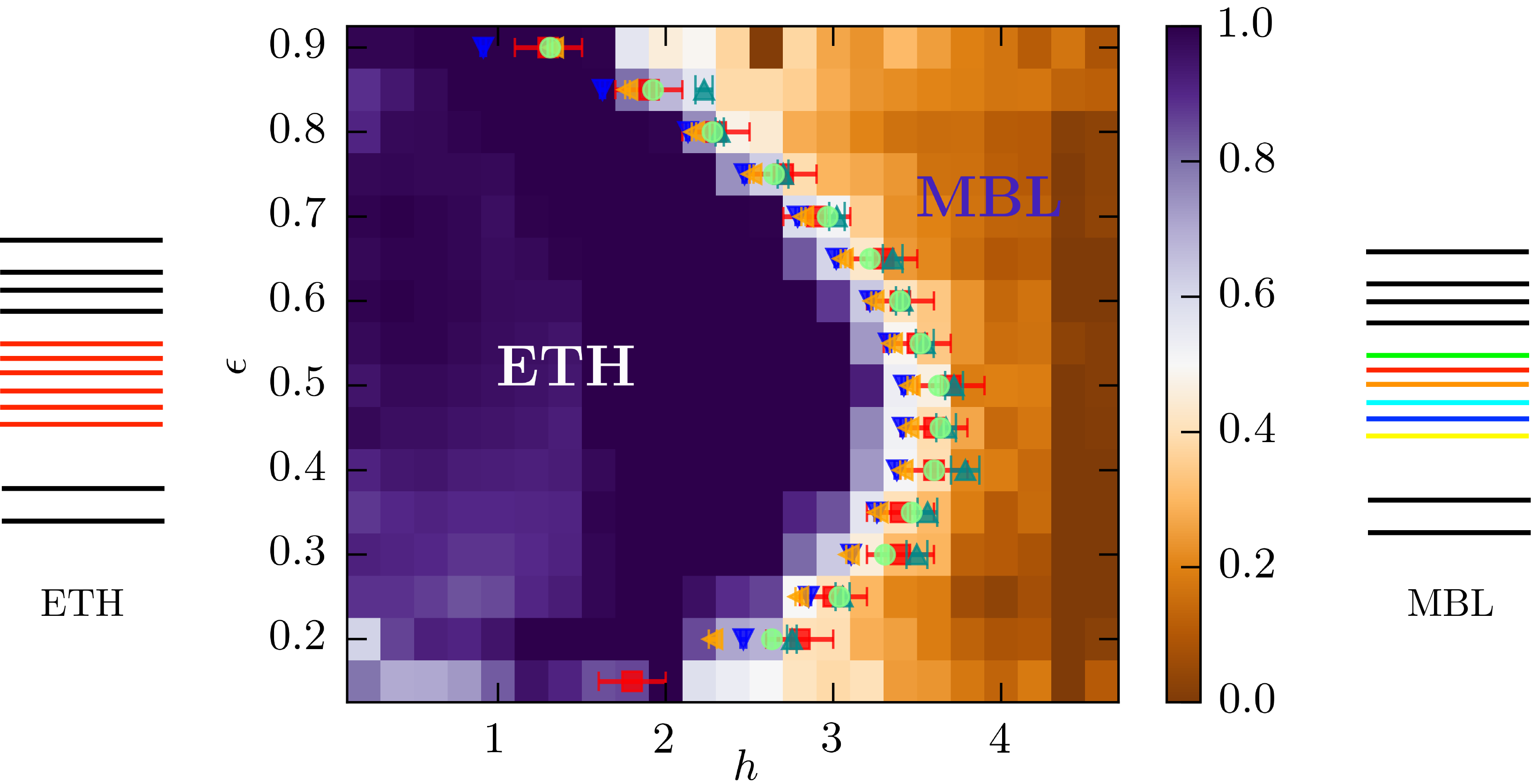}
\caption{Phase diagram of the random field Heisenberg spin chain Eq.~(\ref{eq:rfH}). The vertical axis is the normalized energy density $\epsilon$ of the spin chain ($0 \leq \epsilon \leq 1$), defined as $\epsilon=\frac{E-E_0}{E_m-E_0}$ with $E_0$ ($E_m$) the ground-state (maximum) energy. The horizontal axis is the disorder strength $h$. Left and right: cartoons of the energy spectra in the two phases. Figure extracted and adapted with permission from D. J. Luitz, N. Laflorencie, F. Alet, Phys. Rev. B {\bf 91}, 081103 (2015)~\cite{luitz_many-body_2015}, Copyright (2015) by the American Physical Society.}
\label{fig:PD}
\end{center}
\end{figure}

This model, first introduced in the MBL context in Ref.~\cite{pal_many-body_2010},  is now by far the most studied one, making it a paradigmatic example. Its phase diagram, as obtained numerically in Ref.~\cite{luitz_many-body_2015}, is presented in Fig.~\ref{fig:PD} along with a cartoon representation of the energy levels in the two phases of this phase diagram. This example emphasizes that we will mostly consider in this review (i) {\it isolated} quantum systems (no coupling to a bath), (ii) (eigen-)states at {\it finite energy density} (far from the ground-state, typically in the middle of the spectrum), (iii) lattice systems for which often the energy spectrum is bounded and which are amenable to numerical simulations, albeit not always straightforwardly (see Sec.~\ref{sec:challenge}), (iv) mostly one-dimensional systems (see however Sec.~\ref{sec:dim}) with short-range interactions.

Leaving aside for the moment the details of how this phase diagram was obtained or what the color coding or symbols mean (see Ref.~\cite{luitz_many-body_2015} for this), we start by discussing the two families of eigenstates present in this phase diagram, and the corresponding scenarios: thermal states denoted as ETH (for eigenstate thermalization hypothesis), and MBL (many-body localized) states. We note that other scenarios have been proposed (such as non-ergodic delocalized states~\cite{pino_nonergodic_2016}), but they are likely not relevant for the model Eq.~(\ref{eq:rfH}).

\subsubsection{Thermalization}  
The eigenstates at sufficiently low disorder (left side of Fig.~\ref{fig:PD}) satisfy the eigenstate thermalization hypothesis (ETH)~\cite{deutsch_quantum_1991,srednicki_chaos_1994}, reviewed in details in Ref.~\cite{dalessio_quantum_2016,borgonovi_quantum_2016}. In a nutshell, the ETH makes two assumptions for matrix elements  of few-body (e.g. local) observables ${\rm O}$ in the eigenbasis $|n \rangle$ of an isolated quantum system described by a Hamiltonian $H$:  (i) diagonal matrix elements $\langle n | { O} | n \rangle \simeq {\bar O}(E)$ are smooth functions of the energy E, with ${\bar O}(E)$ matching the microcanonical expectation value of this observable, (ii) off-diagonal matrix elements $\langle n | { O} | m \rangle$ will vanish in the thermodynamic limit, exponentially fast with the microcanonical entropy for the average energy $(E_n+E_m)/2$ (to be more precise, the prefactor will be a smooth observable-dependent function of the energy difference and average, times a random matrix element). These rather strong assumptions were verified numerically in a large variety of systems (see e.g.~\cite{rigol_thermalization_2008,rigol_breakdown_2009,luitz_long_2016}). In practice, we expect ETH to be verified even on small systems for the energy range close to the middle of the spectrum of $H$ where the density of states is important. For the spectrum extrema, convergence to ETH expectations is slower and may not be observed within the range of system sizes where eigenstates can be exactly computed: this is why Fig.~\ref{fig:PD} does not include data for $\epsilon \lesssim 0.15$ and $\epsilon \gtrsim 0.9$. 

One may say that ETH eigenstates living at the same energy density (fixed $\epsilon$) are similar (in the sense of ETH point (i) where they have the same expectation value for few-body observables) which is depicted by the use of the same color for mid-spectrum eigenstates in the left cartoon of Fig.~\ref{fig:PD}. In the ETH phase, the reduced density matrix of a subsystem $A$ (say a few sites in the spin chain model Eq.~(\ref{eq:rfH}))  of an eigenstate ${\hat{\rho}}_A (|n \rangle)= {\rm Tr}_B | n \rangle \langle n | $ (with $B$ the complement of $A$) will take a thermal form ${\hat{\rho}}_A (|n \rangle)\propto \exp(- H/T_n)$. The temperature $T_n$ matches the one of the canonical ensemble reproducing the eigenstate energy, $\langle H \rangle_{T_n}=\langle n | H | n \rangle$. In the middle of the spectrum ($\epsilon=0.5$) which corresponds to infinite temperature, eigenstates behave exactly like random vectors (more precisely, eigenvectors of a random matrix with the same symmetry as the Hamiltonian). In one word, the eigenstates in the ETH phase are thermal and statistical mechanics works.

The ETH has important consequences for dynamics, as it indeed resolves the apparent paradox that most isolated quantum systems, initialized in very different states $|\Psi(t=0) \rangle$, reach the same thermal equilibrium state at long times when following their own unitary dynamics $|\Psi(t)\rangle = e^{-iHt} |\Psi(0) \rangle$ (we denote this as a ``quench experiment''). If ETH holds, one easily sees that any specific information encoded in the initial state will be washed out in the long-time limit, as any (few-body) observable $\langle { O} (t \rightarrow \infty) \rangle  \rightarrow {\bar O}(E_0)$ will converge to the same equilibrium thermal value dictated by the energy $E_0=\langle  \Psi(t=0) | H | \Psi(t=0) \rangle$ of the initial state. One can only emphasize that this thermalization is not due to an external bath (the system is isolated in this quench experiment), it is rather one part of the system which serves as a bath for another part.  

\subsubsection{MBL: avoiding thermalization with disorder}
It was realized early on that a (strong enough) disorder can change the ETH picture.  At strong disorder (right side of Fig.~\ref{fig:PD}), the eigenstates are markedly different. For the model Eq.~\ref{eq:rfH}, the Hamiltonian becomes classical in the strong disorder limit $h\rightarrow \infty$ and the eigenstates are simple product states (e.g. $| \! \uparrow \downarrow \downarrow \uparrow \dots \rangle$ in the $\{S^z\}$ basis), which are readily seen not to satisfy the ETH and not to thermalize. For some systems, the non-interacting limit leads to a free particle model in a strong random potential, mapping to the Anderson problem where localization can occur. The absence of thermalization is there intuitively understood as without interactions, there is no mechanism by which a part of the system can exchange energy/information/particles with another part. For other systems, the model with no interactions is not Anderson localized but does nevertheless display localization properties when disordered interactions are present (see e.g. Ref.~\cite{sierant_many-body_2016,lev_many-body_2016}).

Can this non-thermalizing limit survive when branching in interactions between particles? The denomination {\it many-body localization} (MBL) was coined after the influential works of Ref.~\cite{basko_metalinsulator_2006,gornyi_interacting_2005} who reached a positive answer to this  question using perturbative calculations for disordered fermionic systems. We note several important early contributions~\cite{fleishman_interactions_1980,altshuler_quasiparticle_1997}, including by Shepelyansky and coworkers~\cite{shepelyansky_coherent_1994,jacquod_emergence_1997,georgeot_integrability_1998,georgeot_breit-wigner_1997}, who also considered this very problem (in these early works, it was rather thermalization which was sought for, by adding interactions to a strongly disordered system). 

It is now understood that MBL is the prevalent alternative to thermalization for isolated quantum interacting systems in presence of disorder. The MBL eigenstates, even at the same energy density, are different (as illustrated by different colors in the mid-spectrum eigenstates in the right cartoon of Fig.~\ref{fig:PD}): for instance, the expectation value of a spin component at a given site (a very local observable) for a given eigenstate $\langle n |  S^z_i  | n \rangle$ will generically be different from the expectation value in the next eigenstate $\langle n+1 |  S^z_i | n+1 \rangle$. Again in one word, MBL eigenstates are not thermal and a statistical mechanics description in terms of a canonical ensemble does not work. As emphasized in the review~\cite{nandkishore_many-body_2015}, one shall rather consider a single-eigenstate ensemble, a singular limit of the microcanonical ensemble.

\subsection{Comparison of eigenstates and dynamics in the ETH/MBL phases}

Besides checking the assumptions of ETH, we can use in practice different properties of the eigenstates as well as other probes of thermalization to address the existence of ETH/MBL phases. We mention here a few hallmark properties. The review~\cite{luitz_ergodic_2017} provides a more substantial account, while essentially focusing on the ETH phase.

\subsubsection{Spectral and eigenvalues statistics}

In the ETH phase, eigenvalues/states follow the statistics of Random Matrix Theory. For instance, since $H_{\rm rfH}$ has only real matrix elements in the $\{S^z\}$ basis, the eigenvalues should follow the Gaussian Orthogonal Ensemble (GOE) statistics: Wigner semi-circle law for eigenvalues, Wigner-Dyson law for distribution of level spacings $s_n=E_n-E_{n-1}$. To check this explicitly, one needs to define local average quantities (e.g. dividing the level spacing by the average one for this energy density) or 'unfold' the spectrum, which is not always obvious and free of finite-size effects for many-body systems. Oganesyan and Huse~\cite{oganesyan_localization_2007} introduced a useful computational quantifier, the ratio of consecutive level spacings $r_n=\frac{\min(s_n,s_{n-1})}{\max(s_n,s_{n-1})}$, which is free from local averaging effects. Its average value can be computed in the GOE ensemble to be $\langle r \rangle_{\rm GOE} \simeq 0.5307$. On the other hand, in the MBL phase, eigenvalues are non-correlated, and thus follow a Poisson distribution, for which one obtains a different average value $\langle r \rangle_{\rm Poisson} \simeq 2 \ln(2)-1 \simeq 0.386$. Computing the average value of $\langle r \rangle$ (over eigenstates, and disorder realizations) is an easy task that provides one way to distinguish MBL from ETH using level statistics (one of the symbols in Fig.~\ref{fig:PD} corresponds to the finite-size scaling analysis of $\langle r \rangle$)\footnote{One should be careful to compute level spacings in the same symmetry sector, for instance working only with states with total magnetization $S^z=0$ in the spin chain of Eq.~(\ref{eq:rfH}). Mixing several sectors would erroneously lead to Poisson statistics in the ETH phase.}. Note that an intermediate scenario proposed in Ref.~\cite{serbyn_spectral_2016} and based on a mapping to a Brownian motion process suggests intermediate critical statistics to occur in a large part of the ETH regime preceding the MBL transition point. However, this analysis was criticized in a subsequent numerical study~\cite{bertrand_anomalous_2016}.

What about eigenstates? We recall that ETH eigenstates are similar, whereas MBL ones are different. A possible measure is then to compare the distribution of amplitudes of two eigenstates in a given computational basis (say $\{ | i \rangle\}=\{|S^z\rangle\}$, the usual computational basis for the spin chain Eq.~(\ref{eq:rfH})). This can be achieved by defining the occupation of basis state $| i \rangle$ by the eigenstate $|n\rangle$ as $p_i= | \langle n | i \rangle|^2$, and comparing the two distributions ${p_i}$ and ${p'_i}$ for two nearby eigenstates with the Kullback Leibler divergence: ${\rm KL}=-\sum_i p_i \ln \frac{p_i}{p'_i}$ (note that this quantity depends on the basis $\{ | i \rangle\}$). In the GOE ensemble, two eigenstates are similar, giving rise to a finite value $\langle {\rm KL} \rangle_{\rm GOE}=2$, while $\langle {\rm KL} \rangle$ diverges when increasing system size for MBL eigenstates which become increasingly different.

\subsubsection{Localization in Fock or configuration space?} The notion of occupation of basis state $p_i$ is also useful to try to draw an analogy between the Anderson localization (in real space) and MBL, which is often ascribed as a localization in the Fock/configuration space. This issue will be discussed in more details in Sec.~\ref{sec:randomgraphs}. Let us nevertheless mention already a few aspects of this problem here.

Precursor works by Altshuler {\it et al.}~\cite{altshuler_quasiparticle_1997}, Gornyi {\it et al.}~\cite{gornyi_interacting_2005}, and Basko {\it et al.}~\cite{basko_metalinsulator_2006} showing the existence of a finite temperature metal-insulator transition, were based on the idea of Anderson localization in the Fock space spanned by many-body states represented as Slater determinants built on single-particle (non-interacting) localized states. However, as later discussed by Bauer and Nayak in Ref.~\cite{bauer_area_2013} for a lattice model of 1D interacting fermions in a random potential, building on the simple observation that the connectivity of the Fock space grows {\it{extensively}} with the physical system size, it was concluded that a true Anderson localization is not possible, even deep in the MBL phase. Extensive numerical investigations reached similar conclusions in Refs.~\cite{luca_ergodicity_2013,luitz_many-body_2015} for the random field Heisenberg chain Eq.~(\ref{eq:rfH}) studied in the computational basis (${ | i \rangle}={|S^z\rangle}$) through the participation (Shannon) entropy (measuring how much a given eigenstate extends over the configuration space) $S_{\rm Sh}=-\sum_ip_i\ln p_i$. Such results~\cite{luitz_many-body_2015} are displayed in the color coding of Fig.~\ref{fig:PD} which corresponds to the prefactor (between 0 and 1) of the volume-law scaling of $S_{\rm Sh}$ (see Ref.~\cite{luitz_many-body_2015} for more details). It shows that ETH features full delocalisation while the MBL regime is clearly non-ergodic while not genuinely localized (see Sec.~\ref{sec:randomgraphs}).

\subsubsection{Entanglement properties}
As entanglement is a key notion in our understanding of MBL, we discuss in more details various entanglement aspects in the focus Sec.~\ref{sec:ent_transition} and Sec.~\ref{sec:ent}, and just mention here the main distinction between entanglement in ETH and MBL states: the scaling of their entanglement entropy (EE) with subsystem size. We recall the definition of EE of a pure state $| n\rangle$: we consider a bipartition of the physical system $A\cup B$ in two parts $A$ and $B$ (say $L_A$ and $L_B$ sites of a spin chain of total size $L=L_A+L_B$) and first trace over the degrees of freedom of $B$ to obtain the reduced density matrix ${\hat{\rho}}_A={\rm{Tr}}_B |n\rangle\langle n|$. The eigenvalues of ${\hat{\rho}}_A$ defines the entanglement spectrum $\{\lambda_i\}$, and the von Neumann EE for this bipartition is defined as $S(|n\rangle)=-{\rm{Tr}}\left({\hat{\rho}}_A\ln {\hat{\rho}}_A\right)$.

As already mentioned, ${\hat{\rho}}_A$ of a highly excited eigenstate in the ETH phase can be viewed as a thermal density matrix at finite temperature. Therefore, the corresponding EE is very close to the thermodynamic entropy of the subsystem at finite temperature, thus exhibiting an {\it extensive} ({\it volume-law}) scaling with the subsystem size (scaling as $L_A$ in the spin chain example).
Indeed, eigenstates in the middle of the spectrum are akin to random states for which Page has derived the following scaling for a subsystem $A$ with $L_A$ spins $1/2$: $S \propto L_A\ln2$~\cite{page_average_1993}. At finite energy density $\epsilon >0$, one expects (for sufficiently large $L_A$ and $L_B$) that $S_A = s(E) L_A$ where $s(E)$ is the (thermodynamic) entropy density at the energy corresponding to the eigenstate considered. Such volume-law entanglement at high energy has been indeed observed in the ETH regime of disordered lattice systems~\cite{bauer_area_2013,kjall_many-body_2014,luitz_many-body_2015}. 

\begin{figure}[t!]
\begin{center}
\includegraphics[width=.65\columnwidth,clip]{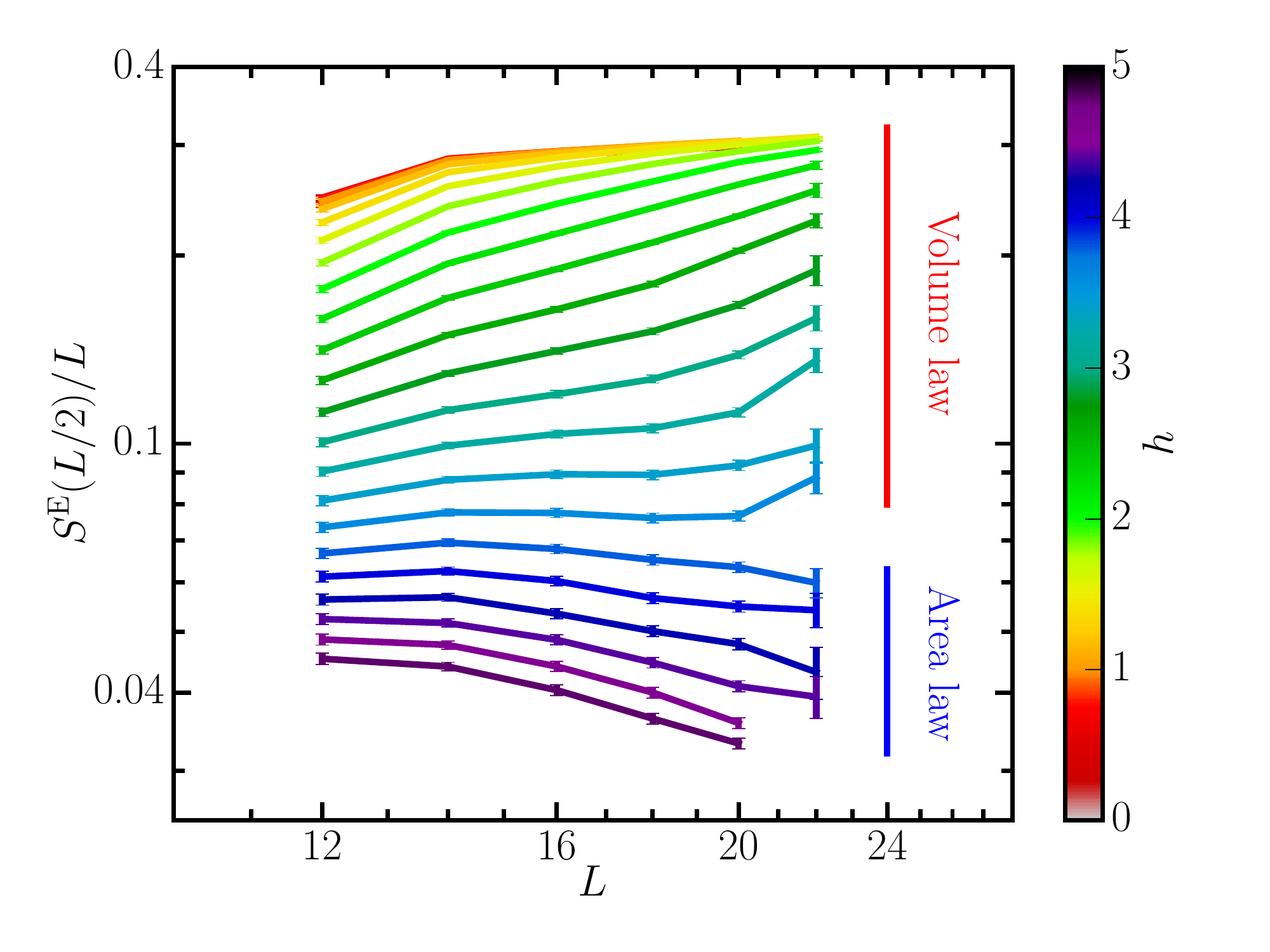}
\caption{Exact diagonalization results for the disorder-average entanglement entropy computed at half-chains $L/2$ for the random field Heisenberg chain model Eq.~(\ref{eq:rfH}) at a fixed energy density in the middle of the many-body spectrum ($\epsilon=0.5$) with $L=12,\ldots, 22$. One sees a qualitative change from volume-law at small disorder $h$ to an area law at $h>h_c\simeq 3.8$.
Figure adapted with permission from D. J. Luitz, N. Laflorencie, F. Alet, Phys. Rev. B {\bf 91}, 081103 (2015)~\cite{luitz_many-body_2015}. Copyright (2015) by the American Physical Society.}
\label{fig:volumearea}
\end{center}
\end{figure}

The MBL eigenstates have radically lower EE. As first observed for interacting fermionic chains in a random potential~\cite{bauer_area_2013}, the scaling of bipartite entanglement entropies is clearly sub-volumic in the MBL regime. Such a small ({\it area-law}) entanglement ($S_A = O(1)$ for a spin chain) was later confirmed in several numerical studies~\cite{kjall_many-body_2014,luitz_many-body_2015,lim_many-body_2016,khemani_critical_2017}. In Fig.~\ref{fig:volumearea} exact diagonalization results on the $S=1/2$ random field Heisenberg chain model Eq.~(\ref{eq:rfH}) are shown for the disorder-average EE of half-chains $L_A=L/2$ for many-body eigenstates at very high energy ($\epsilon=0.5$ in Fig.~\ref{fig:PD}) as a function of size $L$ and disorder strength $h$.  When normalized by the system size, the transition from volume-law to area-law scaling is clearly visible around $h_c\simeq 3.8$. One of the symbols in Fig.~\ref{fig:PD} corresponds to the volume versus area law distinction for the scaling of EE. Note that in the MBL regime, Ref.~\cite{bauer_area_2013} reported a small logarithmic violation of the area law for the maximum entropy obtained from the optimal cut. 
Several important observations are in order: (i) MBL states at finite energy density are related (admittedly in a loose sense at this stage) to {\it ground-states} of many-body systems, for which an area law is generically observed~\cite{eisert_area_2010}. (ii) As they host low entanglement, MBL states can be efficiently represented by matrix product states (MPS) in one dimension, providing a very useful numerical arsenal for capturing highly-excited states, something which is usually not possible (see Sec.~\ref{sec:challenge}). (iii) It was actually proposed that a proper definition of a MBL state is that one can find a finite-depth unitary quantum circuit (with local unitary transformation) that will transform it into a product state~\cite{bauer_area_2013}. 

Let us finally mention that bipartite fluctuations, which are believed to capture entanglement properties for several correlated ground-states~\cite{song_bipartite_2012}, also provides a quantitative tool to track area versus volume law behaviors~\cite{luitz_many-body_2015,singh_signatures_2016}.

 \subsubsection{Transport} 
 The ETH phase is characterized by the transport of energy and conserved quantities (such as the $S^z$ spin component for Eq.~\ref{eq:rfH}), whereas the MBL phase shows absence of transport. 

The spin transport was studied for Eq.~\ref{eq:rfH} in a series of work reviewed in Ref.~\cite{prelovsek_density_2017,luitz_ergodic_2017} through the numerical evaluation of the conductivity $\sigma(\omega)=\frac{1-e^{-\beta \omega}}{2 \omega}\Re\left[\int_{-\infty}^\infty dt e^{i\omega t} \langle J(t) J(0) \rangle\right]$ and its dc part $\sigma_{dc}=\lim_{\omega\rightarrow 0} \sigma(\omega)$ with the spin current operator $J=\frac{1}{L} \sum_i  (S^x_iS^y_{i+1} - S^x_{i+1}S^y_{i})$. In the MBL phase, the dc conductivity was indeed found to vanish $\sigma_{dc}=0$, for any finite $T$. In the ETH phase at very low disorder, one expects on general ground to find a metallic behavior with  $\sigma_{dc} > 0$, even though this is very difficult to show numerically. The detailed nature of the transport is a subject of current discussion, with a possibility of subdiffusive behavior near the transition to the MBL phase, instead of the expected diffusive behavior (see Sec.~\ref{sec:sub}). 

\subsubsection{Dynamics after a quench} 
\label{sec:dyn}
Following the dynamics after a quench from some initial state $| \Psi(0) \rangle$ also provides a sharp distinction between the two phases. 
In the ETH phase, one observes a fast transfer of information through the system. At long times the system reaches thermal equilibrium, with a temperature associated to the initial energy of the state, as discussed above. This quantum information transfer  is characterized by a rapid growth of EE of the time-evolved state $ | \Psi(t) \rangle = e^{-iHt} | \Psi(0) \rangle$, which is ballistic $S_A(|\Psi(t) \rangle) \propto t$~\cite{kim_ballistic_2013} (in first approximation, see discussion in Sec.~\ref{sec:sub} for the impact of subdiffusion on this behavior) -- eventually the EE will saturate to its equilibrium value $s(E_0) L_A$. Equivalently, any local observable $\langle \Psi(t) | O | \Psi(t) \rangle$ will equilibrate towards its canonical ensemble expectation value in the limit of long times. The memory of the initial state is lost: consider for instance initializing the N\'eel state $| \Psi(0) \rangle=| \!\uparrow \downarrow \uparrow \downarrow \dots \rangle$ which has a maximal staggered magnetization $m_s = \frac{2}{L}\sum_j (-)^j S_j^z=1$. At long times, it will equilibrate to  $\langle \Psi(t\rightarrow \infty) | m_s | \Psi(t\rightarrow \infty) \rangle=0$, its thermodynamic expectation value.

On the other hand, the MBL phase keeps memory of its initial state. For instance, $\langle \Psi(t\rightarrow \infty ) | m_s | \Psi(t\rightarrow \infty) \rangle\rightarrow m* > 0$ (see e.g.~\cite{ros_remanent_2016}), and in general all local observables will equilibrate at long times, but not to the canonical ensemble~\cite{serbyn_quantum_2014}. Another symbol in Fig.~\ref{fig:PD} is a quantifier of the memory of an initial state (even though not obtained in a quench setup). 

A signature of the MBL phase is that, even in the absence of transport of energy or spin/particles, quantum information spreads, albeit at a slow rate: the EE after a quench grows logarithmically with time $S_A(|\Psi(t) \rangle) \propto \ln(t)$~\cite{znidaric_many-body_2008,bardarson_unbounded_2012,serbyn_universal_2013,vosk_many-body_2013,andraschko_purification_2014}. This is a specific trait of MBL that marks a difference with Anderson insulators which also show no transport, but for which $S_A(|\Psi(t) \rangle)$ is bounded. Other ways of distinguishing an Anderson from a MBL insulator are discussed in Sec.~\ref{sec:distinguish}. The entropy will eventually saturate in the long-time limit to a value which scales with the size $L_A$, but for which the prefactor is {\it lower} than the thermal value $s(E_0)$. The origin of this log growth is discussed in Sec.~\ref{sec:liom}. 

\subsection{Many-body mobility edge}
\label{sec:edge}
At large enough disorder, all states are localized in what is often referred to as the full-MBL (fMBL) phase. The phase diagram of Fig.~\ref{fig:PD} presents however a feature not yet discussed: the location of the transition line between MBL and ETH states appears to vary with the energy density. This is a so-called many-body mobility edge, in analogy with the mobility edge present for the Anderson transition. Several other numerical studies beyond Ref.~\cite{luitz_many-body_2015} also find a mobility edge in the same or other models (e.g.~\cite{kjall_many-body_2014,baygan_many-body_2015,laumann_many-body_2014,mondragon-shem_many-body_2015,villalonga_exploring_2018}). The importance of a mobility edge at finite energy {\it density} was emphasized in the early work of Basko, Aleiner and Altshuler~\cite{basko_metalinsulator_2006}, as it implies a finite transition temperature  $T_c>0$, below which MBL physics set in and transport vanish.

The existence of a mobility edge has however been debated on theoretical grounds. The main line of argument is based on the effect of a local fluctuation which has a higher energy than the background (a 'hot bubble'), has thermal properties, is furthermore mobile and can thus thermalize the other 'cold' parts of the sample (see Ref.~\cite{de_roeck_absence_2015} for more details). This type of fluctuation is always possible, and the authors of Ref.~\cite{de_roeck_absence_2015} argue that no many-body mobility edge can exist in the thermodynamic limit. In this view, computations which see a mobility edge are not performed on large enough system sizes to see this asymptotic regime, a scenario which is hard to test with numerics.

\subsection{Capturing the essence of the MBL phase: local integral of motions}
\label{sec:liom}

Of utmost importance to the understanding of the MBL phase is the fact that it has a simple description in terms of quasi-local degrees of freedom which commute with the Hamiltonian: they are referred to as l-bits (for localized bits) or lioms  (local integrals of motions). They were first discussed in a phenomenological description of the MBL phase~\cite{huse_phenomenology_2014,serbyn_local_2013}, and can now be constructed mathematically in a rigorous fashion for a class of spin chains~\cite{imbrie_many-body_2016,imbrie_diagonalization_2016}. Excellent reviews on the lioms approach to MBL can be found in Ref.~\cite{imbrie_review:_2016,rademaker_many-body_2017}, we extract here only the main ideas.

At each site $i$ of the system (say the spin chain of Eq.~\ref{eq:rfH}), a Pauli-spin~\footnote{To stick to the literature, we use Pauli instead of spins $1/2$ operators here (connection with Eq.~\ref{eq:rfH} is ${S}^\alpha=\frac{1}{2} \sigma^\alpha$, $\tilde{h}_i =h_i/2$)} operator $\tau^z_i$ (a l-bit or liom), is constructed using only the original spin degrees of freedom on this site and 'nearby' sites: $\tau^z_i = \sigma^z_i + \sum_{j,k} \sum_{\alpha,\beta=x,y,z} c^{\alpha,\beta}(i,j,k) \sigma^\alpha_j \sigma^\beta_k + \ldots$ with coefficients $c$ that decay exponentially with the distance between $i$ and other sites $j,k$ (hence quasi-locality). The key feature is that there exists such a quasi-local transformation which completely diagonalizes the Hamiltonian: 
\begin{equation}
\label{eq:Hliom}
H=-\sum_i \tilde{h_i} \tau_i^z + \sum_{i,j} J_{i,j} \tau_i^z \tau_j^z+ \sum_{i,j,k} J_{i,j,k} \tau_i^z \tau_j^z \tau_k^z + \dots. 
\end{equation}

The resulting Hamiltonian is completely {\it classical}: it does not imply l-bit flip operators $\tau^{x,y}$. All $\tau^z$'s at all sites commute with each other, and with the Hamiltonian, hence they form a complete set of lioms. The first term is a local field term (in the strong disorder limit, one identifies $\tilde{h}_i = {h_i}/2$ for Eq.~(\ref{eq:rfH})), and all further terms $J_{i,j,...}$ are interactions between the lioms, which also decay exponentially with the distance between lioms. These interaction terms are crucial, as they distinguish a MBL from an Anderson insulator (for which all $J_{i,j,...}$ strictly vanish), and have important dynamical consequences.

The fact that the many-body Hamiltonian can be diagonalized solely using a (quasi-)local unitary transform is quite remarkable, as it is {\it almost never} the case for generic many-body systems: it signals some emerging local integrability of the MBL phase. There are many different ways to construct this transformation: by perturbation from the strong disorder limit, by considering infinite-time expectation values of local observables etc (see e.g. Ref.~\cite{serbyn_local_2013,chandran_constructing_2015,monthus_many-body_2016,rademaker_explicit_2016,imbrie_many-body_2016,monthus_many-body-localization_2017,goihl_construction_2017,kulshreshtha_behaviour_2017,mierzejewski_counting_2017,scardicchio_perturbation_2017}). Connections between the perturbative approach of Basko, Aleiner and Altshuler~\cite{basko_metalinsulator_2006} used for lioms and the original lioms phenomenology was performed in Ref.~\cite{ros_integrals_2015}. We note the beautiful analogy between the construction of lioms and the Fermi liquid approach (which also diagonalizes an interacting Hamiltonian into a quasi-particle basis)~\cite{bera_many-body_2015,bera_one-particle_2017,rademaker_many-body_2017,lezama_fate_2017}.

From this picture of lioms, one can recover all hallmarks of the MBL phase. A many-body eigenstate is simply given by the set of eigenvalues $\pm 1$ for each $\tau^z_i$, e.g. $| 1 ,1, -1, 1, ... \rangle_\tau$. 'Static' properties of the MBL such as the area law for entanglement entropy (which is proportional to the number of lioms 'cut' by the boundary between the two subsystems), Poisson statistics, different local observables for nearby eigenstates are direct consequences of this simple labelling. 
The memory of the initial state after a quench, as well as the fact that all local observables equilibrate to a given finite value at long-times are also deduced from the overlap of such observables with the conserved operators $\tau^z_i$.

What about dynamics towards these equilibration values? This is where the interactions between lioms intervene: they provide a mechanism for {\it dephasing}. Interactions induce correction to the simple precession around the z-axis due to the first term $\tilde{h}$, and albeit at a slow rate, are able to entangle two spins at any distance: the low growth of entanglement (characteristic of MBL) is related to the exponentially weak interactions between the lioms. The precise way of how all local observables decay to the long-time values are discussed in Ref.~\cite{serbyn_quantum_2014} and in Sec.~\ref{sec:distinguish} as one way to distinguish MBL from Anderson localization.

The existence of lioms is now seen by many authors as {\it the} defining property of the MBL phase. For instance, analytical arguments regarding possible existence of MBL in dimension larger than 1 (Sec.~\ref{sec:dim}), or in presence of certain symmetries (Sec.~\ref{sec:sym}), equate presence of MBL with existence of lioms. There are however some aspects of MBL physics which, at the moment, are not described in terms of lioms. For instance, there is no direct lioms picture for the existence of a many-body mobility edge (see Sec.\ref{sec:edge}) --models for lioms imply that all states are localized--, or for what happens at the MBL phase transition.  We again invite readers interested in this integrability structure of the MBL phase to consult the existing reviews~\cite{imbrie_review:_2016,rademaker_many-body_2017}.

\subsection{Lattice models for MBL}
\label{sec:models}

Besides the random-field Heisenberg spin chain Eq.~\ref{eq:rfH}, several other lattice models have been used to investigate MBL, mostly in a one-dimensional geometry. We cite only a few here, without mentioning all possible variants (in lattice geometry, range or type of coupling constants etc) or discussing systems with long-range~\cite{burin_energy_2006,hauke_many-body_2015, burin_many-body_2015,nandkishore_many_2017} or mean-field interactions~\cite{laumann_many-body_2014,baldwin_many-body_2016,burin_localization_2016,ponte_thermal_2017,georgeot_integrability_1998,georgeot_quantum_2000}.

The closely related $S=1/2$ XXZ spin chain in a random field
\begin{equation}
\label{eq:XXZ}
H_{\rm XXZ} = \sum_i S_i^x S_{i+1}^x + S_i^y S_{i+1}^y + \Delta S_i^z S_{i+1}^z -\sum_i h_i S_i^z
\end{equation}
is a generalization of Eq.~\ref{eq:rfH} (obtained at $\Delta=1$) to anisotropic interactions. It is  equivalent via a Jordan-Wigner transformation to the interacting spinless fermionic model 
\begin{equation}
\label{eq:fermions}
H_{\rm sf} =  \frac{1}{2} \sum_i c_i^\dagger c^{\vphantom{\dagger}}_{i+1} + c^{\vphantom{\dagger}}_i c_{i+1}^\dagger +  V \sum_i (n_i-1/2) (n_{i+1}-1/2) -\sum_i \mu_i n_i
\end{equation}
with $V=\Delta$, $\mu_i=h_i$, $c^\dagger_i,c_i$ creation and annihilation operators, $n_i=c_i^\dagger c_i$ occupation operator at site $i$ (open boundary conditions are assumed to make a strict correspondance with $H_{\rm XXZ}$). Both models are useful to make a connection with the Anderson transition, as the non-interacting case $V=\Delta=0$ corresponds to a free-particle model in a random potential. These models also have a U(1) symmetry, corresponding to the conservation of total spin magnetization $S^z$ (total charge number $\sum_i n_i$).

The nature of the disorder is encoded in the distribution of $h_i$: the most-studied distributions are the box distribution ($h_i$ uniformly distributed in $[-h,h]$), the bimodal distribution $h_i = h$ ($-h$) with a probability $p$ ($1-p$)~\cite{tang_quantum_2015,andraschko_purification_2014}, and the quasi-periodic case $h_i= h \cos (2 \beta \pi i)$ with $\beta$ an irrational number~\cite{iyer_many-body_2013,naldesi_detecting_2016,setiawan_transport_2017}. The latter case corresponds when $V=\Delta=0$ to the well-studied Aubry-Andr\'e model~\cite{aubry_analyticity_1980}. All these distributions lead to a MBL phase at large disorder strength $h$.

Another spin model has also been studied: the Ising chain in a transverse field (with next-neighbor interactions)
\begin{equation}
\label{eq:Ising}
H_{\rm Ising}  = \sum_i J_i^1 S_i^z S_{i+1}^z +   \sum_i J_i^2 S_i^z S_{i+2}^z - \sum_i \Gamma_i S_i^x.
\end{equation}
It has a smaller $Z_2$ (Ising) symmetry, with a conserved total parity $P=\prod_i S_i^x$. The case with no next-neighbor interactions $J_2=0$ can also be mapped to a free-fermion model (alternatively a different term such as $\sum_i J_i^2 S_i^x S_{i+1}^x$ can be chosen for the same purpose). The disorder can be put either in the Ising $J^{1,2}$  or transverse field $\Gamma$ terms. These models also show MBL behavior at large disorder, with the addition of a new feature: inside the MBL phase, the $Z_2$ symmetry can be broken or not, resulting in two different MBL phases (see Sec.~\ref{sec:Ising}).

Another important model with respect to experimental perspectives (Sec.~\ref{sec:exp}) is the spinfull fermionic 1d Hubbard model 
\begin{equation}
\label{eq:Hubbard}
H = -t \sum_{i,\sigma=\uparrow,\downarrow} c_{i,\sigma}^\dagger c_{i+1,\sigma}^{\vphantom{\dagger}} + c_{i,\sigma}^{\vphantom{\dagger}} c_{i+1,\sigma}^\dagger
+U \sum_i n_{i}^\uparrow n_{i}^\downarrow-\sum_i \mu_i (n_{i}^\downarrow+n_{i}^\uparrow)
\end{equation}
with $c^\dagger_{i,\sigma},c_{i,\sigma}^{\vphantom{\dagger}}$ creation and annihilation operators, $n_i^\sigma=c_{i,\sigma}^\dagger c_{i,\sigma}^{\vphantom{\dagger}}$ occupation operator of a fermion with spin $\sigma$ at site $i$, with a random (e.g. $\mu_i$ taken in a box distribution) or quasi-periodic $(\mu_i=\Delta \cos (2 \beta \pi i)$) potential. MBL features have been found at large disorder, albeit with an important observation~\cite{prelovsek_absence_2016}: the charge sector appears localized (with for instance a non-vanishing charge density correlation  $\langle m(0)m(t) \rangle$ at long-time with $m=\sum_i (-)^i (n_i^\uparrow + n_i^\downarrow)$) while the spin sector shows no localization (the spin density correlator $\langle s(0)s(t) \rangle$, with $s=\sum_i (-)^i (n_i^\uparrow - n_i^\downarrow)$,  vanishes at long time). This absence of a true fully localized phase could be related to the $SU(2)$ symmetry of this model (see discussion in Sec.~\ref{sec:sym}). Indeed, when different disorders $\mu_i^\uparrow n_i^\uparrow, \mu_i^\downarrow n_i^\downarrow$ are chosen for different spin orientations (random magnetic field breaking the SU(2) symmetry), full MBL is recovered~\cite{prelovsek_absence_2016}. The explicit construction of independent lioms for the Hubbard model confirms this scenario: the total number of lioms with a support on $M$ consecutive sites is found to be $4^M-1$ for the random magnetic field case (thus full-MBL) and to be smaller $~ 4^M /2$, albeit interestingly quite large, when the disorder does not couple to spin~\cite{mierzejewski_counting_2017}.

We also note that the 1d bosonic Hubbard model~\cite{sierant_many-body_2016},  the related 1d model of Josephson Junction arrays~\cite{pino_multifractal_2017}, and the 1d $t$-$J$ model~\cite{lemut_complete_2017} have been studied with disordered potential or interactions, with also MBL behavior found at large disorder.  

\section{Review of experiments}
\label{sec:exp}

Can the intriguing aspects of MBL developed above be experimentally probed? This is a particularly legitimate question since most of the conclusions on MBL have been reached for {\it closed, isolated} systems, a situation problematic for experiments due to the inevitable coupling to the environment. While initial development of the MBL field was driven by theory (with numerics contributing actively), several major experiments, mostly on cold-atoms, changed the situation. The first series of experiments described below were all based on ultracold atomic and molecular setups (cold atoms, trapped ions) which are reasonably well isolated from baths and decoherence, and for which disorder, and in some cases interactions, are also well controlled. These experiments often follow the dynamics after a quench from a simple initial state. Several questions are opened by these new developments which we briefly review here.

\begin{figure}
\begin{center}
\includegraphics[width=0.6\columnwidth]{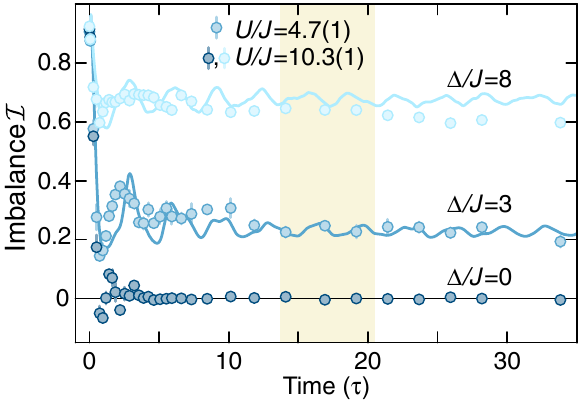}
\caption{Experimental measurement of the time-dependence of the imbalance $I(t)$ for various strengths of the quasi-periodic potential $\Delta$ (and interactions $U$). Figure taken from M. Schreiber {\it et al.}, Science {\bf 349}, 842 (2015)~\cite{schreiber_observation_2015}. Reprinted with permission from AAAS.}
\label{fig:Imb}
\end{center}
\end{figure}

A first cold-atom experiment showing a signature of a MBL (localized) phase was presented in the group of I. Bloch, with the study of the post-quench dynamics of a charge density wave (selected as an initial state and consisting of an almost perfect alternation of filled and empty sites) in a 1d ultracold fermionic gas with a quasi-periodic potential~\cite{schreiber_observation_2015}. The system is well described by the Hubbard model with a quasi-periodic potential, of strength denoted by $\Delta$, see Sec.~\ref{sec:models} and Eq.~(\ref{eq:Hubbard}). Interactions are controlled with the help of Feshbach resonances, while the quasi-periodic potential is obtained by superposing two laser beams with incommensurate frequencies. Fig.~\ref{fig:Imb} displays as a function of time (measured in hopping units) the experimentally measured imbalance, which is the normalized difference between the number of particules in odd ($N_{\rm odd}$) or even ($N_{\rm even}$) sites of the optical super-lattice: $I(t) = \frac{|N_{\rm odd}(t)-N_{\rm even}(t)|}{N_{\rm odd}(t)+N_{\rm even}(t)}$. Measurements of site occupation is performed with band-mapping techniques~\cite{schreiber_observation_2015}.

The initial maximal imbalance rapidly decays to its equilibrium zero value in absence of the quasi-periodic potential. The MBL phase is on the other hand characterized by a non-vanishing value of the imbalance at long time, meaning that the system keeps a memory of the initial state (as encoded in such a local measurement). The same group~\cite{bordia_coupling_2015} finds absence of MBL in a two-dimensional system of coupled identical one-dimensional tubes (with the same quasiperiodic potential as in the previous work).

Another experiment on chains of trapped ions~\cite{smith_many-body_2016} performed by the group of C. Monroe, can be modeled by an Ising model with long-range interactions in a random transverse field. The effective spin 1/2 degrees of freedom correspond to hyperfine states of $^{171}$Yb$^+$ ions. Long-range Ising interactions (with a tunable power-law exponent ranging from $\alpha=0$ to 3) are generated by stimulated Raman transitions whereas the disorder in the form of a random transverse field can be 'programmed' experimentally using ac Stark shifts.  Using this setup, Ref.~\cite{smith_many-body_2016} finds clear signatures of a MBL phase, as probed by a non-zero value of staggered magnetization in the long-time after a quench from an initially programmed N\'eel state (this measure is identical to imbalance in the cold-atom experiments above). Recently, a similar experiment~\cite{zhang_observation_2017} was used to create a time crystal by driving the MBL phase with an oscillating field (see Ref.~\cite{hess_non-thermalization_2017} for a review of such experiments).

These experiments very nicely confirmed the theoretical prediction that MBL is possible for 1d lattice systems, in presence of strong disorder. Further recent experiments are now able to address questions that are  extremely challenging for theory: study of the phase transition, of dynamics in the thermal phase in 1d and in the full phase diagram of two-dimensional systems.

Using a quantum microscope setup with interacting bosonic atoms in a two-dimensional optical lattice with a disordered potential, Ref.~\cite{choi_exploring_2016} found signatures similar to those in previous 1d fermionic experiments (non-zero saturation value of imbalance), pointing towards the existence of a MBL phase in 2d. At slight variance with the 1d case, in this experiment the imbalance was measured with respect to an initial state where half of the system was full of particles, and the other half empty. The role of interactions as well as the continuous nature of the transition have been also addressed in this work. The fermionic experimental setup has also been adapted to simulate a two-dimensional system with independent quasi-periodic potential in both directions~\cite{bordia_probing_2017}, finding at large disorder almost no relaxation for imbalance --pointing again towards 2d MBL-- as well as a striking slowing down of imbalance dynamics at intermediate disorder (before the transition). The typical system sizes probed experimentally in two-dimensions are by several orders of magnitude larger than the (very limited) ones accessible to numerics in two-dimensions, fully following the logic of analog quantum simulators. Future prospects include reaching even larger systems. 

Even for 1d systems where theoretical tools are more advanced, experiments can probe regimes or physics that are difficult to study from a theoretical view point. Using the 1d quasi-periodic setup, L\"uschen {\it et al.}~\cite{luschen_signatures_2016} studied the influence of laser-induced controlled dissipation, showing that the imbalance decays in a strechted-exponential form. The impact of a periodic drive (with a modulated amplitude of the quasi-periodic potential $\Delta(t)= \Delta + A \sin( \omega t)$) on the MBL phase observed in 1d was studied in Ref.~\cite{bordia_periodically_2017}. For a strong drive $A=\Delta$, experiments indicate that the MBL phase found at large $\Delta$ can be destroyed and replaced by a drive-induced ergodic phase at sufficiently slow frequency. Finally, a slowing down of transport was also observed in the thermal phase as approaching the transition~\cite{luschen_observation_2017}.

Let us now conclude by mentioning a few different experimental situations where MBL has been considered or discussed. Systems with dipolar interactions such as ultra-cold polar molecules or NV centers in diamonds have been discussed in the context of MBL~\cite{yao_many-body_2014}. The later case of NV centers was experimentally investigated using 'black' diamond (with many centers) where a slow subdiffusive (power-law) relaxation was reported~\cite{kucsko_critical_2016}: whether MBL is (partly) at play here is an open issue. In the different context of condensed matter physics where, instead of response after a quench, transport properties are usually probed, it is important to mention that a dramatic drop of conductivity at low temperature in disordered indium-oxide films (close to a zero temperature superconducting-insulator transition driven by a magnetic field) has been interpreted as the signature of entrance into a MBL phase~\cite{ovadia_evidence_2015}. A recent NMR-type experiment in a nuclear spin chain (fluorapatite) has also been argued in favour of a MBL phase in such a system~\cite{wei_exploring_2018}. Connections between MBL physics and the dynamical nuclear polarization process have also been highlighted, with a polarization maximum close to the MBL transition~\cite{de_luca_dynamic_2015,de_luca_thermalization_2016,rodriguez-arias_exactly_2017}.

\section{Focus on a selection of topics}
\label{sec:focus}

\subsection{MBL and symmetries}
\label{sec:sym}

The existence of phases in equilibrium statistical mechanics is traditionally related to the notion of spontaneous symmetry breaking in the standard Landau approach, with a local order parameter characterizing the broken symmetry. More recent uses of symmetries beyond the Landau picture include the classification of topological matter and various types of short/long-range entangled phases. It is legitimate to also ask about the role of symmetries for the MBL phase, which is not adequately captured by equilibrium statistical mechanics. Probably the full picture of what types of MBL phase(s) and phase transition(s) are compatible with a given symmetry is not present yet; we nevertheless provide two important examples of the relation between symmetries and MBL.

\subsubsection{Order in MBL states: Ising and topological}
\label{sec:Ising}

One example where the MBL phase evades several of the important paradigms of equilibrium statistical mechanics is the existence of long-range order in one-dimensional systems (with short-range interactions) at finite energy density -- which counterpart in equilibrium at finite temperature is forbidden~\cite{PhysRev.187.732}. Counterintuitively, disorder may help in ordering. This was discussed first in Refs.~\cite{huse_localization-protected_2013,bahri_localization_2015,bauer_area_2013} and pedagogically reviewed in Ref.~\cite{nandkishore_many-body_2015}, from which we sketch the basic idea through the following example: consider a lattice spin model (e.g. Eq.~\ref{eq:Ising}) which has an Ising symmetry characterised by a parity operator $P$, even in presence of disorder. 
Each eigenstate $|n,p \rangle$ of the system on finite samples has a fixed parity eigenvalue $p=\pm 1$, but this symmetry can be broken spontaneously in the thermodynamic limit. On finite large systems indeed, there is a phase for which the states $|\phi_n^\pm \rangle=  \frac{1}{\sqrt{2}}(|n,p \rangle \pm |n,-p \rangle)$ (which break the parity symmetry) can have an exponentially long life-time. This is completely similar to the more familiar ferromagnetic states $| \! \uparrow \uparrow \uparrow \ldots \rangle$ and $ | \! \downarrow \downarrow \downarrow \ldots \rangle$ in a pure Ising quantum chain (with no disorder): they are {\it not} the finite-size ground-states in the ferromagnetic phase --- rather the ground-states are close to the parity preserving states $ \frac{1}{\sqrt{2}}(| \uparrow \uparrow \uparrow \ldots \rangle \pm  | \downarrow \downarrow \downarrow \ldots \rangle)$ which are extremely close in energy for strong Ising coupling ---, but they have an exponentially long life-time (with the size of the chain) and eventually become the symmetry-broken ground-states in the thermodynamic limit.

Why a similar phenomenon can occur in the MBL phase for excited states?  The physical idea is that domain walls, which usually destroy the long-range Ising order at finite temperature in equilibrium statistical mechanics, are now localized even at finite energy density~\cite{huse_localization-protected_2013}. The long-range order, which is not of ferromagnetic but rather of spin-glass type, is preserved in excited states and can be captured by considering adapted dynamical spin-spin correlations functions as well as a spin-glass-like susceptibility~\cite{kjall_many-body_2014,pekker_hilbert-glass_2014}. Thus inside the phase diagram of a disordered model with such a symmetry, one can find both a 'regular' paramagnetic MBL phase (with no order present in the excited states) and a 'spin-glass' MBL phase (with order present). This later phase can additionally present a pairing of eigenvalues on finite-size spectra~\cite{huse_localization-protected_2013} (similar to the pairing of the two $p=\pm 1$ ground-states in the broken symmetry phase of the pure Ising model). These two types of MBL phases have been observed in numerical simulations~\cite{pekker_hilbert-glass_2014,kjall_many-body_2014,vosk_dynamical_2014}. 

The nature of the phase transition between these two MBL phases has been addressed in a few works~\cite{pekker_hilbert-glass_2014,vosk_dynamical_2014,vasseur_quantum_2015,monthus_random_2018} using in particular strong disorder renormalization group techniques adapted to excited states~\cite{pekker_hilbert-glass_2014}. This transition hosts a new type of MBL-critical states, which could form a full phase under certain conditions for a type of disordered critical models: this is the  'quantum critical glass'  (see Sec.~\ref{sec:glass}). Note also that for certain models, the MBL phase can be shown~\cite{potter_symmetry_2016,vasseur_particle-hole_2016,friedman_localization-protected_2017,prakash_ergodicity_2017} to {\it necessarily} break such a symmetry (and be a MBL-spin glass), in line with the idea that not all symmetries are compatible with MBL (see Sec.~\ref{sec:glass}). 

The same type of argument can be used and extended to show the existence of topological order in excited MBL states for a family of models, for instance in Ref.~\cite{bahri_localization_2015,chandran_many-body_2014} where signatures of $Z_2 \times Z_2$ symmetry protected order are found in highly excited states. The compatibility of the MBL phase with various types of order (including symmetry protected topological order) has been also discussed in Ref.~\cite{huse_localization-protected_2013,slagle_many-body_2015,potter_protection_2015} (see also Sec.~\ref{sec:glass}). Overall, all these possibilities are in line with the idea that MBL states are good platforms to host new phases of matter, as they are immune to thermalization and do not allow transport of excitations that usually destroy these phases at equilibrium.

\subsubsection{MBL-compatible symmetries} 
\label{sec:glass}
Here we summarize a recent work by Potter and Vasseur~\cite{potter_symmetry_2016} (see also Ref.~\cite{protopopov_effect_2017}) who argued that MBL (in the strict sense of Sec.\ref{sec:liom} with exponentially localized lioms) is not possible for systems which have a non-Abelian symmetry. The argument is based on the following observation: the global symmetry is promoted to a local one in an (exponentially localized lioms) MBL phase. This creates an exponential number of degeneracies for each energy level, as there are $p>1$ extra-quantum numbers needed to label each liom for non-Abelian symmetries. Resonances between different energy levels will be strongly favored by this large degeneracy (i.e. even a small perturbation will have the corresponding matrix elements), resulting in the destruction of MBL.

According to Ref.~\cite{potter_symmetry_2016}, possible ways out for non-Abelian symmetries are thus the following:  (i) a MBL phase happens but it is accompanied by a simultaneous breaking of the larger symmetry group to a smaller Abelian group, (ii) a quantum critical glass (for non-Abelian groups with finite number of irreducible representations), which is a critical-form of MBL states which are neither thermal nor localized (again in the sense of exponentially localized lioms), (iii) or, probably more ubiquitously, thermalization and no MBL at all. 
The arguments presented in this work start with the definition of MBL as given by the existence of local integral of motions (Sec.~\ref{sec:liom}) -- the authors argue however that the general results, based on symmetry arguments, should even hold for (putative) MBL states which are not described by lioms.

These results complement those obtained on the compatibility of MBL with symmetry protected topological order~\cite{slagle_many-body_2015,potter_protection_2015}. Case (i) above has been observed in numerical simulations, leading to rich phase diagrams~\cite{vasseur_particle-hole_2016,friedman_localization-protected_2017}. The quantum critical glass (case (ii)) has been predicted by a real-space renormalization group method to occur in non-Abelian spin chains~\cite{vasseur_quantum_2015}. These new forms of critical MBL states have specific signatures (see the reviews Ref.~\cite{parameswaran_eigenstate_2017,vasseur_nonequilibrium_2016}): critical correlations in eigenstates, log correction to the area law, as well as modified forms (powers of log) of entanglement dynamics after a quench. In a sense, they are quantum critical points promoted to exist in each high-energy excited state. Quantum critical glasses have not been observed in microscopic models yet.

\subsection{The thermal-MBL transition}
\label{sec:transition}

\subsubsection{Generalities}
The nature of the phase transition between the thermal / ergodic phase and the MBL phase is quite subtle and elusive, and let us already emphasize that it continues to defy a full theoretical understanding. We only briefly review some of the approaches attempted, but the interested reader is referred to a recent review~\cite{parameswaran_eigenstate_2017} which describes in more detail several aspects discussed below. 

First, let us mention what this phase transition is not. Since it separates a regular 'thermal' phase from a non-thermodynamic (MBL) phase, the accumulated knowledge on equilibrium statistical mechanics of phase transitions is not always helpful. It is often described as an 'eigenstate phase transition', since an exponential number of eigenstates (located at the same energy density) suddenly change properties in a radical way, as well as a 'dynamical phase transition', as it leaves no trace in standard thermodynamics -- only dynamical probes are expected to see the  transition. There is no simple mean-field or Landau theory for this transition, or no simple limit or microscopic toy model where it is understood (see however Ref.~\cite{monthus_many_2016,monthus_finite_2016}). Also, most theoretical approaches are restricted to one-dimensional lattice systems, which are so far the only ones (together with some mean-field models) where the MBL phase is ascertained. 

The current consensus is that, in one dimension, the transition to the MBL phase has both continuous and discontinuous aspects triggered by non-local effects. This conclusion is reached from finite-size numerics and phenomenological renormalization group (RG) approaches. Numerics which address the transition are mostly obtained with exact diagonalization on modest sample sizes (at most 22 spins~\cite{luitz_many-body_2015} for the model Eq.~(\ref{eq:rfH}), which nevertheless corresponds to a large configuration space of dimension $7.10^5$), are consistent with finite-size scaling form of continuous transitions, which a correlation / localization length which diverges as a power-law of the distance to the critical point: $\xi \sim |h-h_c|^{-\nu}$. Most numerical studies~\cite{luitz_many-body_2015,kjall_many-body_2014} obtain an exponent $\nu\sim 1$, which violates a bound~\cite{chandran_finite_2015} $\nu \geq 2/d$, an equivalent of the Harris bound for stability of continuous phase transitions with respect to disorder. This discrepancy is often attributed to numerics not being in the correct scaling regime (because of too small system sizes), even though the very applicability of this bound has been discussed~\cite{monthus_many_2016}. 

On the other hand, two phenomenological models for the transition have originally been proposed~\cite{vosk_theory_2015,potter_universal_2015}, based on some ad hoc strong disorder renormalization group rules. Both have in common to consider regions (clusters) of MBL and thermal clusters which can merge according to some renormalization rules, depending on their internal level splitting as well as tunnelling amplitudes between clusters (see discussion in Sec.~\ref{sec:dim}). The two approaches differ by the precise renormalization rules and the fact that Ref.~\cite{potter_universal_2015} finds that the transition is driven by only a backbone of thermal regions with long-range entanglement, which are very diluted (they occup a finite fraction of the sample at the transition), but act as an effective bath for the rest of the system composed of MBL regions. This  approach has been further improved~\cite{dumitrescu_scaling_2017} by providing a more precise description of the thermal blocks, making connection with the picture of ergodic grains discussed in Ref.~\cite{de_roeck_many-body_2017}. Both RG schemes find a continuous phase transition with a critical exponent $\nu  \sim 3$, consistent with the bound. 

Let us finally note that a partly different renormalization scheme has been recently proposed~\cite{thiery_many-body_2017,thiery_microscopically_2017}, with an analysis of the resonant couplings pointing towards an avalanche mechanism, where an ergodic block of small size is able to thermalize an arbitrary large localized sample as soon as the localization length in this region is large enough. The outcome of this analysis~\cite{thiery_many-body_2017,thiery_microscopically_2017} is the distinction between a non-diverging typical localization length and an average localization length which diverges from the MBL side, but stays finite in the thermal side up to the transition point -- hence showing again both continuous and discontinuous aspects. A numerical test of the validity of the avalanche scenario has been performed in a spin chain model (with ergodic grains represented by random matrices) in Ref.~\cite{luitz_how_2017}.

\subsubsection{Entanglement properties}
\label{sec:ent_transition}
The key role and specific behavior of entanglement at the transition has been highlighted in most approaches. Numerics see a peak of the variance of entanglement at the transition~\cite{kjall_many-body_2014,luitz_many-body_2015},  and entanglement is one of the key parameters in the RG descriptions~\cite{vosk_theory_2015,dumitrescu_scaling_2017}. 

\paragraph{Critical entanglement entropy}The precise behavior of the scaling of entanglement entropy right at the transition is still debated. This question was first addressed by Grover in Ref.~\cite{grover_certain_2014} who considered the case of a small subsystem $L_A\ll L$. Assuming the entropy to be a continuous scaling function of $L_A/\xi$ (where $\xi$ is a characteristic length scale diverging at the transition) with no $L$-dependence, the strong subadditivity of entanglement led to the fact that $S(L_A)$ must display a thermal volume-law scaling at the MBL criticality. Exact diagonalization studies~\cite{luitz_many-body_2015,devakul_early_2015,lim_many-body_2016} are compatible with a sub-thermal volume-law entanglement entropy at criticality, while not in the relevant regime of small subsystem $L_A\ll L$. In another limit, which does not addresses the  scaling regime $L\gg L_A\gg 1$, Khemani {\it{et al.}}~\cite{khemani_critical_2017} focused on the $L_A=1$ case (one spin) for which a strong sub thermal entropy was found, suggesting a discontinuity of $S$ at the ETH-MBL transition. A similar conclusion was reached in Ref.~\cite{dumitrescu_scaling_2017}, based on an improved renormalization group~\cite{potter_universal_2015}. While not exact, this technique allows to deal with very large system sizes and to be in the scaling regime $L\gg L_A\gg 1$. For system sizes $L\gg\xi$, the short interval entanglement displays a discontinuous jump across the transition from fully thermal volume-law on the ETH side, to an area-law on the MBL side. An alternative approach, developed by Monthus in Ref.~\cite{monthus_many-body-localization_2016} for a toy-model having a MBL critical point very close to Poisson statistics, concluded for an intermediate critical scaling $S(L_A)\sim L_A^\alpha$ with $0<\alpha<1$, neither volume nor area law. The discrepancy between these two results~\cite{dumitrescu_scaling_2017,monthus_many-body-localization_2016} could be attributed to the different level statistics inherent to these two approaches.

\begin{figure}[ht!]
\begin{center}
\includegraphics[width=\columnwidth,clip]{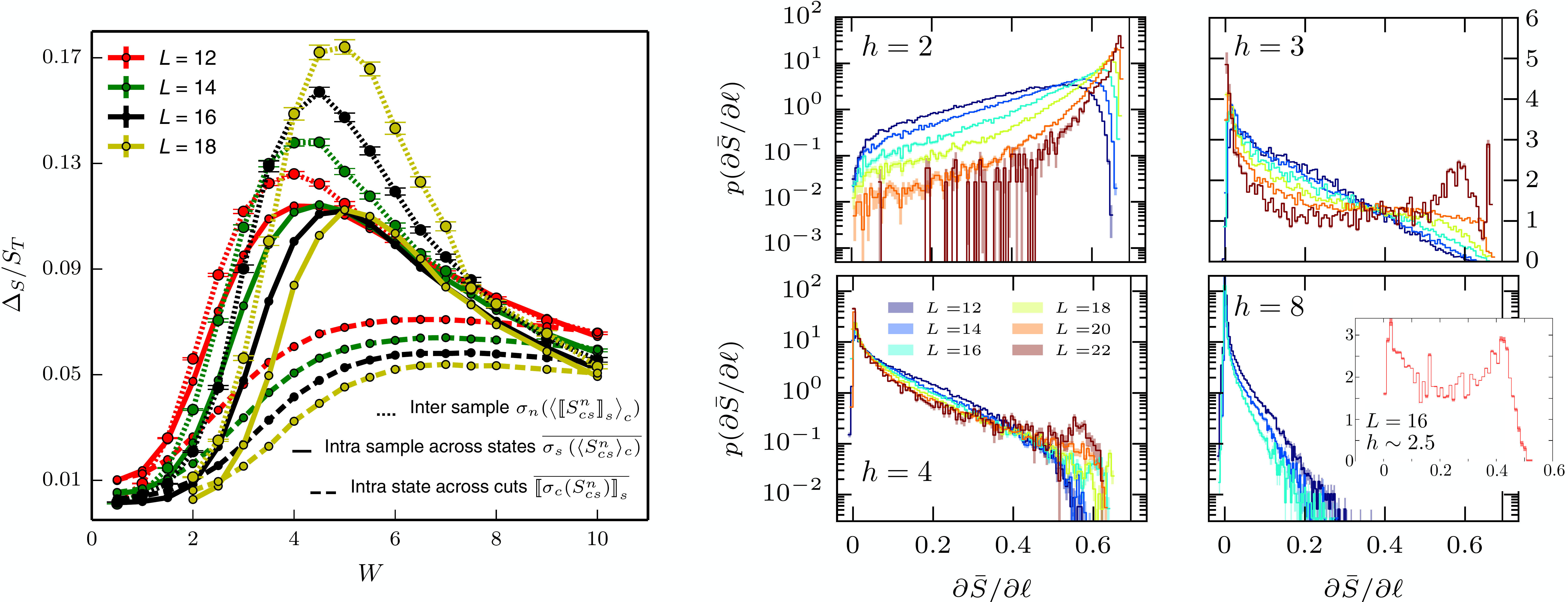}
\caption{Numerical results illustrating the peculiar properties of the EE distributions across the ETH-MBL transition. Right: from Yu {\it{et al.}}~\cite{yu_bimodal_2016} for model Eq.~(\ref{eq:rfH}), by increasing the disorder $h$ the distributions of entanglement slopes (computed among different samples and eigenstates in the middle of the spectrum) display a clear qualitative change. When the transition is approached, a bimodal structure is identified. This structure is also visible for a single random sample (inset) for 6000 eigenstates. Figure extracted with permission from X. Yu, D. J. Luitz, B. K. Clark, Phys. Rev. B {\bf 94}, 184202 (2016)~\cite{yu_bimodal_2016}, copyright (2016) by the American Physical Society. Left: from Khemani {\it{et al.}}~\cite{khemani_critical_2017}, the standard deviation $\Delta_S$ (normalized by the thermal entropy $S_T$) is shown as a function of disorder strength $W$ for model Eq.~(\ref{eq:rfH}) with additional second neighbor $xy$ coupling (yielding a critical disorder $W_c\sim 7$). Depending on the set over which $\Delta_S$ is computed (across samples, cuts, eigenstates), the variance displays different properties. Figure extracted from V. Khemani, S. Lim, D. Sheng, D. A. Huse, Phys. Rev. X {\bf 7}, 021013 (2017)~\cite{khemani_critical_2017}.}
\label{fig:S_variance}
\end{center}
\end{figure}

 \paragraph{Critical distributions} 
Instead of simply looking at the disorder/eigenstate average entropies, a closer look into their distributions is very instructive, as first discussed in Refs.~\cite{bauer_area_2013,kjall_many-body_2014,luitz_many-body_2015,lim_nature_2016} where an enhancement of the variance was reported when approaching the ETH-MBL transition. 
Results from Yu {\it{et al.}}~\cite{yu_bimodal_2016} for model Eq.~(\ref{eq:rfH}) are discussed in Fig.~\ref{fig:S_variance} (right),  where the four small panels show the qualitative change in the distributions of entanglement slopes (which quantifies the volume-law prefactor, averaged over all cuts in each chain). In this work, a bimodal structure for such distributions was shown to develop at criticality, a feature which is quite surprisingly also observed at the level of a {\it single} disorder realization (see inset, where the distribution is computed from eigenstates in the {\it{same}} disorder sample). As argued in two recent studies~\cite{khemani_critical_2017,khemani_two_2017}, a key for understanding the ETH-MBL transition lies in the differences between fluctuations of entanglement coming from different eigenstates in the {\it same} disorder sample, from those coming from different samples (see Fig.~\ref{fig:S_variance} taken from Ref.~\cite{khemani_critical_2017}).

Ref.~\cite{khemani_two_2017} speculates that there are accordingly two different universality classes for the MBL/thermal transition, one dominated by eigenstate fluctuations (in the same sample), and a second one by the quench disorder. On too small system sizes, the second universality class could be hidden by the first one, for which the bound $\nu \geq  2/d$ (derived for quench disorder) does not apply. This scenario is consistent with numerical simulations for a spin chain with quasi-periodic disorder, a case which is argued to have milder finite-size effects. The transport properties of such quasi-periodic chains have also been investigated in this light~\cite{setiawan_transport_2017}. In any case, the ultimate critical behavior for systems with a true random potential is still unknown.

\subsubsection{Possible directions} What are the prospects for a better understanding of the transition? Regarding theory and numerics in particular, this is intimately linked to progresses in computational methods for MBL (see Sec.~\ref{sec:challenge} for a more detailed account of the available methods). There is some hope that advances in matrix-product states or operator techniques could be pushed to be able to capture eigenstates closer to the phase transition, coming from the MBL side. The important caveat there is that these variational methods are usually biased towards low-entanglement states, which can be problematic if the distribution of entanglement of eigenstates is broad near the transition as observed in exact diagonalization~\cite{luitz_long_2016,yu_bimodal_2016} (see Sec.~\ref{sec:ent_transition}). Another possibility, which also relies on specific properties of the MBL phase, is to improve diagonalization by a flow of unitary transforms. 

Starting from the other direction, one can maybe get insight from standard numerical techniques (e.g. quantum Monte Carlo or series expansion) which are valid on this thermalizing side of the transition, e.g. by tracking diverging signatures of transport properties as one approaches the transition. Two new recent important advances in the family of matrix-product states methods, namely using a time-dependent variational principle~\cite{leviatan_quantum_2017} or a density matrix truncation method~\cite{white_quantum_2018}, show great promises for simulating dynamics also in the thermal phase, giving hope to capture the transition.

And finally, experiments in the near future will probably be able to address in more details and/or complement theory on several properties of this phase transition. 

\subsection{Entanglement spectroscopy}
\label{sec:ent}
Besides the scaling of entanglement entropy, and given the success of this approach for pure many-body problems~\cite{li_entanglement_2008,laflorencie_quantum_2016}, several works~\cite{yang_two-component_2015,geraedts_many-body_2016,serbyn_power-law_2016,pietracaprina_entanglement_2017,gray_many-body_2017,geraedts_characterizing_2017,yang_entanglement_2017} considered the information contained in the entanglement spectrum (ES) $\{\lambda_i\}$ of the reduced density matrix (RDM) ${\hat{\rho}}_A$ of eigenstates or time-evolved states. One often writes the RDM as ${\hat{\rho}}_A=e^{-H_E}$ with $H_E$ the entanglement Hamiltonian, with eigenvalues $-\ln (\lambda_i)$ forming a 'pseudo-energy'  spectrum.

\subsubsection{Entanglement spectrum properties for the MBL problem}

\begin{figure}[t!]
\begin{center}
\includegraphics[width=.8\columnwidth,clip]{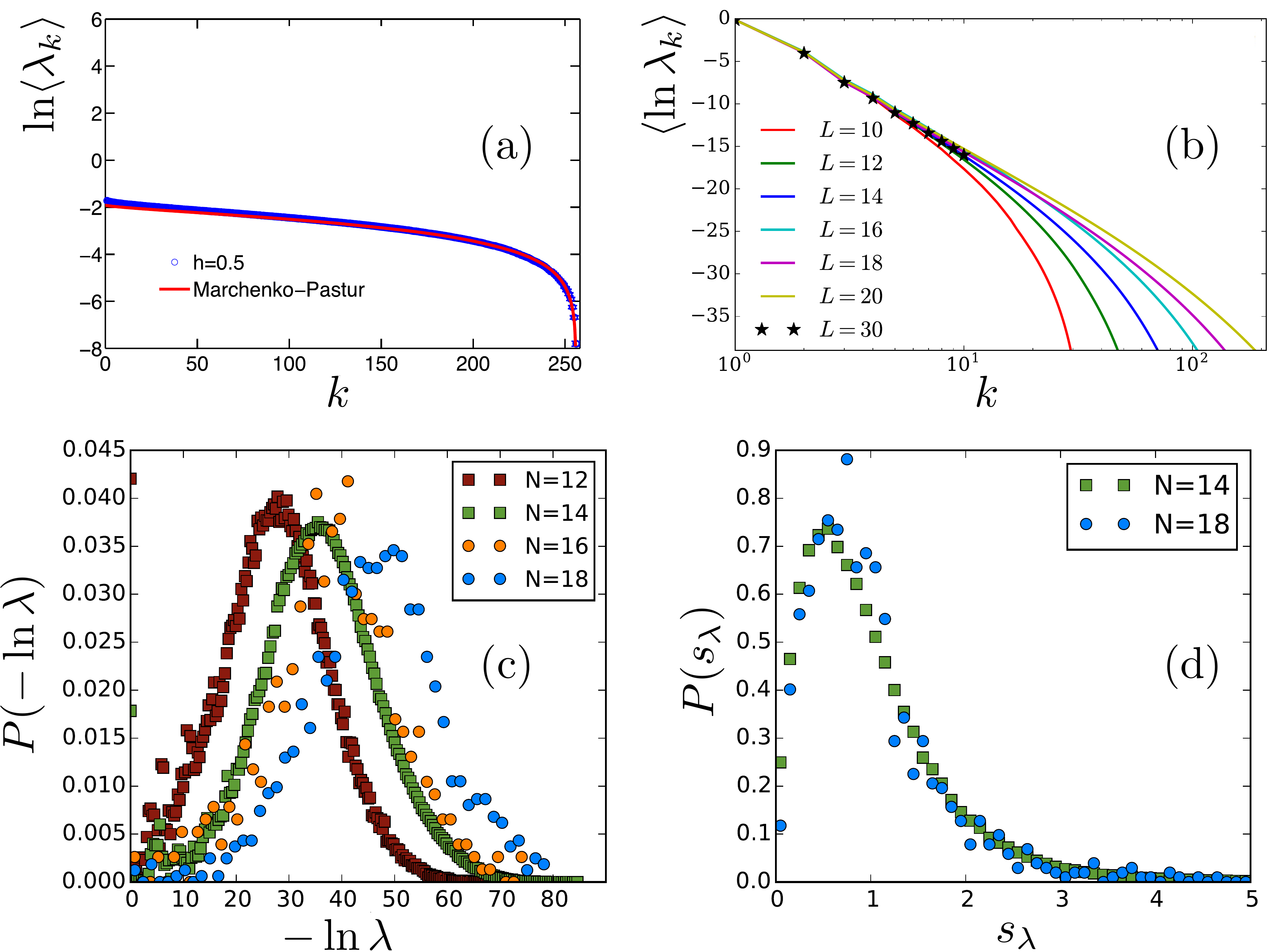}
\caption{Entanglement spectra for various situations computed using exact diagonalization (except (b) $L=30$ obtained with MPS) for cuts at half-chain. (a) From Yang {\it{et al.}}~\cite{yang_two-component_2015}, in the ergodic regime the (disorder average) entanglement levels computed for $L=16$ are well  described by the Mar\v{c}enko-Pastur distribution (see text). Figure extracted with permission from Z.-C. Yang {\it et al.}, Phys. Rev. Lett. {\bf 115}, 267206 (2015)~\cite{yang_two-component_2015}, Copyright (2015) by the American Physical Society. (b) From Serbyn {\it{et al.}}~\cite{serbyn_power-law_2016}, in the MBL regime the first sequence of entanglement levels follows a power-law. Figure extracted with permission from M. Serbyn {\it et al.}, Phys. Rev. Lett. {\bf 117}, 160601 (2016)~\cite{serbyn_power-law_2016}, Copyright (2016) by the American Physical Society.  (c-d) From Geraedts {\it{et al.}}~\cite{geraedts_many-body_2016}, in the MBL regime, the high-energy part of the ES whose density is shown in panel (c) displays a semi-Poisson distribution for the level spacing $s_\lambda$ (d). Figures extracted with permission from S. D. Geraedts, R. Nandkishore, N. Regnault, Phys. Rev. B {\bf 93}, 174202 (2016)~\cite{geraedts_many-body_2016}, Copyright (2016) by the American Physical Society.}
\label{fig:es}
\end{center}
\end{figure}
First in the ergodic regime, the RDM spectrum is very well described by the Mar\v{c}enko-Pastur (MP) distribution, which characterizes the eigenvalues density of a random Wishart matrix~\cite{marcenko_distribution_1967}, as shown in Refs.~\cite{yang_two-component_2015,pietracaprina_entanglement_2017}, see Fig.~\ref{fig:es} (a).  In this thermal regime, the entanglement level statistics is also well described by the random matrix theory of appropriate ensemble (GOE or GUE depending on the symmetry of the Hamiltonian)~\cite{geraedts_many-body_2016}. Interestingly, the MP form remains quite robust in the 'high-energy' part of $H_E$ (smallest entanglement levels) when approaching the MBL transition while the 'low-energy' part starts to deviate. This is in contrast with usual expectations for ground-states of either pure~\cite{li_entanglement_2008,calabrese_entanglement_2008}  or disordered systems~\cite{leiman_correspondence_2015} for which the universal part is rather expected in the 'low-energy' sector. 

In the MBL phase, the distribution of entanglement levels qualitatively changes from the flatness of the MP distribution towards a more singular form, as shown in Ref.~\cite{serbyn_power-law_2016}. Interestingly, a clear two-component structure of the ES shows up in the MBL regime. The first entanglement levels $\lambda_1\ge \lambda_2\ldots$ obey a power-law distribution $\lambda_i\sim 1/i^\gamma$~\cite{serbyn_power-law_2016} for the low-energy part, as visible in Fig.~\ref{fig:es} (b), with an exponent $\gamma\propto 1/\xi$ ($\xi$ being a characteristic many-body localization length). This specific behavior has important consequences for numerics based on matrix-product states (see Sec.~\ref{sec:challenge}) as the number of retained entanglement levels required to match the exact EE (within a given precision) is quite low in the MBL phase, in contrast with the ETH phase where this number grows with system size (see e.g. Fig. 4 in Ref.~\cite{serbyn_power-law_2016}).

The high-energy part, while representing a totally subdominant weight in the MPS-like description of MBL eigenstates, seems to retain some universal information as emphasized in Ref.~\cite{geraedts_many-body_2016} where a level statistics of the semi-Poisson form was found for the distribution of ES gaps $s_{\lambda_i}=-\ln\lambda_{i+1}+\ln\lambda_i$, with level repulsion (in contrast with the true energy spectrum, which follows a genuine uncorrelated Poisson distribution), see Fig.~\ref{fig:es} (c-d). In addition let us also mention the analytical result of Monthus~\cite{monthus_many-body-localization_2016} who discussed the multifractal properties of the ES in the MBL regime and at the transition.

\subsubsection{Dynamics of the entanglement spectrum}
\begin{figure}[t!]
\begin{center}
\includegraphics[width=\columnwidth,clip]{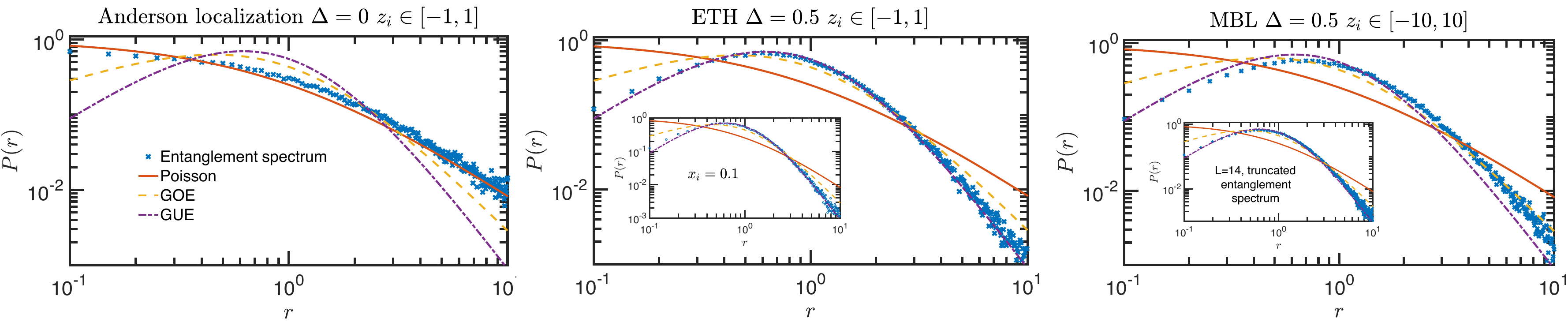}
\caption{From Yang {\it{et al.}}~\cite{yang_entanglement_2017}, exact diagonalization results obtained for $L=12$ (unless specified) Heisenberg $S=1/2$ chains with i.i.d. random fields in longitudinal $z_i$ and transverse $x_i$ directions. Entanglement is computed from $L/2$ bipartitions. The distribution of entanglement gap ratio $P(r)$ with $r_i=(\lambda_i-\lambda_{i-1})/(\lambda_{i+1}-\lambda_{i})$ at time $t=1000$ (in units of hopping) after a quench from a non-entangled initial product state is displayed in various situations. Left: non-interacting Anderson localization where the ES displays Poisson statistics. Center: ETH regime showing GUE behavior. Right: MBL regime with GUE statistics emerging if the largest entanglement levels are discarded (inset). Figures extracted with permission from Z.-C. Yang {\it et al.}, Phys. Rev. B {\bf 96}, 020408 (2017)~\cite{yang_entanglement_2017}, Copyright (2017) by the American Physical Society.}
\label{fig:es_t}
\end{center}
\end{figure}

So far, we discussed entanglement spectroscopy of eigenstates. One can also learn interesting physics using out-of-equilibrium protocols, such as the one proposed by Yang {\it{et al.}}~\cite{yang_entanglement_2017}. In this work, whose main results are displayed in Fig.~\ref{fig:es_t}, the level statistics of the ES has been studied through the distribution of the gap ratio $P(r)$ with $r_i=(\lambda_i-\lambda_{i-1})/(\lambda_{i+1}-\lambda_{i})$ at time $t=1000$ (in units of hopping) after a quench from a non-entangled initial product state. While for non-interacting Anderson localization (Fig.~\ref{fig:es_t} left) uncorrelated Poisson statistics emerge for the entanglement levels after a long time, the situation is qualitatively different when interactions are present.
In this case, the ES after a long time $t=1000$ displays Wigner-Dyson (of GUE type) statistics for both ETH and MBL regimes (Fig.~\ref{fig:es_t} center and right panels). This strongly contrasts with the energy level statistics which signals a qualitative difference between  Wigner-Dyson in the ETH regime {\it{vs.}} Poisson for the MBL phase, but is not able to discriminate between Anderson and MBL insulators (both showing a Poisson behavior).  More precisely in the MBL regime, the ES statistics approaches Wigner-Dyson slowly in time $\sim 1/\ln t$~\cite{yang_entanglement_2017}, a behavior interpreted in light of the lioms (Sec.~\ref{sec:liom}) which mutual interaction decays exponentially with their relative distance. 

\subsection{Role of disorder}
\label{sec:disorder}
\subsubsection{Type of disorder. true-random versus quasiperiodic, bond versus sites}

In the model discussion of Sec.~\ref{sec:models}, we introduced several types of disorder terms: true random disorder (box or bimodal distribution, we can also consider Gaussian distribution) versus quasi-periodic potential; disorder which acts on kinetic or potential terms (identically one can distinguish bond versus site disorder) etc. While they all have in common the possibility of promoting a MBL phase, we briefly mention a few important differences. 

True disorder admit the possibility of Griffiths-regions, regions of space where the disorder is anomalously different (say much larger) than average. These regions have been argued to be responsible for the subdiffusive regime observed in the {\it thermal} phase in several 1d models~\cite{agarwal_anomalous_2015}, as they act as bottlenecks for transports. Quasi-periodic potentials do not allow for such regions (they may nevertheless show slow subdiffusion from other mechanisms). 

Regarding the transition (see Sec.~\ref{sec:transition}): first, the bound derived~\cite{chandran_finite_2015} (an extension of the result by Chayes {\it et al.}~\cite{chayes_finite-size_1986}) for the correlation length exponent is $\nu \geq 2/d$ for the true random disorder, stronger than the bound expected from the Harris-Luck criterion~\cite{luck_classification_1993} (but not derived) for a quasi-periodic potential $\nu \geq 1/d$. Second, as discussed in Sec.~\ref{sec:transition}, the quasi-periodic and true random disorder may have different finite-size effects close to the transition, on top of (possibly) different universality classes. 

We also note the under-exploited possibility to use binary disorder (e.g. $h_i=\pm h$) as it offers the unique possibility to perform an exact average over disorder using series expansion techniques~\cite{tang_quantum_2015} or ancillary spin in DMRG-like methods~\cite{andraschko_purification_2014,enss_many-body_2016}. 

Finally, let us make the simple remark that the symmetries which are preserved/broken by the type of chosen disorder are crucial (see Sec.~\ref{sec:sym}) -- as commonly known for the Anderson transition. For instance, for a random quantum spin 1/2 Heisenberg chain, if the disorder in on-site such as the random-field case of Eq.~(\ref{eq:rfH}), the model symmetry is reduced to $U(1)$, in which case a MBL phase is possible at large disorder. On the other hand, if the disorder is solely located on the bonds (random-bond chain), the SU(2) symmetry is preserved and the model has no MBL phase~\cite{vasseur_quantum_2015,protopopov_effect_2017}. 

\subsubsection{Is disorder needed?}

The above question was asked quite early, in the quest for a so-called translation-invariant MBL phase~\cite{carleo_localization_2012,roeck_asymptotic_2014,schiulaz_ideal_2014,de_roeck_scenario_2014,barbiero_out_2015} (see also the precursor work of Ref.~\cite{kagan_localization_1984}). Several ideas emerge with different mechanisms by which an effective disorder is generated in an otherwise clean system. Several of the proposals are based on the mixture of slow and fast particles (where the slow particles effectively act as a disorder potential)~\cite{schiulaz_ideal_2014,schiulaz_dynamics_2015,yao_quasi_2016}, some others are based on quantization of classical glassy models (where the effective disorder is created by a dynamical constraint)~\cite{hickey_signatures_2016,van_horssen_dynamics_2015}. Another idea is to allow a higher dimensional local Hilbert space (e.g. models with spins $S>1/2$) which would be slower to explore and could leave room for slowing-down and MBL~\cite{carleo_localization_2012,pino_nonergodic_2016}.

Most of these proposals were backed up by numerics that showed several hallmarks of MBL: slow (compatible with log) growth of entanglement, slow dynamics after a quench, Poisson-like spectral statistics etc. Diagnosis in terms of entanglement after a partial measurement (e.g. on the slow particles) has also been identified in the closely related proposal of quantum disentangled liquids~\cite{grover_quantum_2014} .

Most of these models were based however on a separation of energy scales in the Hamiltonian: high/low hopping term for the fast/slow particles, low potential to kinetic energy ratio etc. This can become problematic due to the finite samples used in the numerics, which are in a sense less representative of the 'thermodynamic limit'  than for models which do not have such separation of scales. This idea was highlighted in the work by Papic, Stoudenmire and Abanin~\cite{papic_many-body_2015}  which shows that the density of states for several of the proposed models is given by well-separated bands for the system sizes considered in the regime where MBL was 'detected' -- whereas as on larger systems, the bands would eventually merge and MBL hallmarks disappear. In another words, this means that in a given regime of parameters, small systems can mimic the phenomenology of MBL with extremely long transient times before actually reaching thermal equilibrium. This regime dubbed 'quasi-MBL'~\cite{yao_quasi_2016} or 'asymptotic localization'~\cite{roeck_asymptotic_2014} is certainly interesting in its own right and could perhaps be explored by explicitly engineering small systems (such as Josephson Junction arrays) in the good parameter regime. 

The distinction between quasi-MBL and true MBL requires access to very long-time dynamics on large-enough systems, which is not always possible with numerics in all parameter ranges. While accordingly we cannot provide a definitive statement, it appears that most previous proposals of translation-invariant MBL (with possible exceptions of Ref.~\cite{garrison_partial_2017,mondaini_many-body_2017}) show this transient quasi-MBL regime. A recent work rationalizes this slow regime in terms of a perturbation theory which allows to classify the different processes and associated time scales which ultimately lead to delocalization~\cite{michailidis_slow_2018}. Other recent ideas have been explored to circumvent this using again a two-species construction, by e.g. considering Anderson-like localization of free particles in an effective disorder induced by another locally conserved degree of freedom variable~\cite{smith_disorder-free_2017,yarloo_anyonic_2018,smith_absence_2017}.

A bubble-like argument~\cite{de_roeck_scenario_2014} (see Sec.~\ref{sec:edge}, Sec.~\ref{sec:dim}) also predicts that ultimately there is no full translation-invariant MBL: a small resonant bubble will always delocalize and thermalize the full system. Note that whereas the conclusions of this bubble argument seem to be corroborated by the numerics mentioned above, it is not the case for other situations with bubble-arguments, most notably the prediction of absence of mobility edge~\cite{de_roeck_absence_2015} (see Sec.~\ref{sec:edge}). This thus leaves some uncertainty on whether such asymptotic phenomenological arguments are relevant/backed up by simulations on the currently available system sizes.

\subsection{Anomalous transport in the thermal regime}
\label{sec:sub}

Has the existence of a MBL phase or phase transition any reverberation outside of its location in the phase diagram of a disordered system? A large number of works has observed some anomalously slow dynamical properties in several systems (mostly one-dimensional) in the thermal portion of the phase diagram, which is often described in terms of a 'sub-diffusive' regime prior to the transition. This regime is not free of controversies/uncertainties, as can be seen in the very good reviews~\cite{agarwal_rare-region_2017,prelovsek_density_2017,luitz_ergodic_2017} which summarize the current knowledge on this topic.  

The slow dynamics in this 'sub-diffusive' regime has been evidenced (from numerical simulations) by sub-linear decay of the autocorrelation function of the local magnetization/density $\langle S_i^z(0)S_i^z(t)\rangle \propto t^{-\beta}$~\cite{agarwal_anomalous_2015,bar_lev_absence_2015}, power-law decay of imbalance $I(t) \propto t^{-\gamma}$ or growth of entanglement entropy $S(t)\propto t^{1/z_{\rm ent}}$ after a quench~\cite{luitz_extended_2016}, as well as a specific frequency dependence of the ac conductivity $\sigma(\omega) \propto \omega^{1-2/z}$~\cite{agarwal_anomalous_2015,gopalakrishnan_low-frequency_2015}. All the exponents in these expressions do no take their diffusive value, but rather vary continuously with the disorder strength in the thermal regime. Relations between these different exponents should be satisfied under some general conditions, however the numerical evidence for an agreement between exponents is far from being unquestionable~\cite{luitz_ergodic_2017}. On the other hand, an analysis~\cite{nahum_dynamics_2017} based on a coarse-grained surface growth picture suggests that exponents corresponding to various growing length scales associated to different quantities (entanglement, conserved quantities, operator spread etc) should differ. A connection between off-diagonal matrix elements (in the eigenbasis) and subdiffusive exponents has also been pointed out~\cite{luitz_anomalous_2016}. One should also note that other numerical studies of the dc conductivity~\cite{steinigeweg_typicality_2016,barisic_dynamical_2016} conclude on the other hand at a non-zero value in the thermal phase (in contradiction with the subdiffusive scenario~\cite{agarwal_anomalous_2015,gopalakrishnan_low-frequency_2015}), and that other numerics deduce from the analysis of density propagators that the subdiffusive regime is actually transient~\cite{bera_density_2017}.

Besides theory, it is important to note that two recent experiments~\cite{luschen_observation_2017,bordia_probing_2017} also observe a slow decay of imbalance (after a quench) compatible with a power-law decay with a varying exponent (with additional log corrections in Ref.~\cite{bordia_probing_2017}), at least on the time scales accessible before background decays become too important.

The extent of this subdiffusive part of the thermal regime has also been debated~\cite{agarwal_anomalous_2015,luitz_extended_2016,khait_transport_2016,steinigeweg_typicality_2016,barisic_dynamical_2016} in the model of Eq.~(\ref{eq:rfH}): is the transport always sub-diffusive as soon as disorder is present, or a transition between diffusive / sub-diffusive regimes exist? The most advanced numerical work~\cite{znidaric_diffusive_2016} on large spin chains (using a different technique where the transport is probed in a spin chain weakly coupled to a boundary bath) conclude that there is a purely diffusive phase at weak disorder in this model, favoring the second scenario. 

Assuming that there is slow transport inside the thermal phase (at least on the sizes/time scales probed by numerics and experiments), its origin is not fully understood yet. The most prominent scenario~\cite{agarwal_anomalous_2015} is the existence of spatial regions where the disorder is anomalously high, which act as bottlenecks for transport in one-dimensional systems. One can show that indeed these so-called Griffiths regions (reviewed in Ref.~\cite{agarwal_rare-region_2017}) in the thermal phase do create slow transport in one-dimension, with varying power-laws, as they result in a broad distribution of relaxation times. The phenomenological renormalization group treatments of the transition~\cite{vosk_theory_2015,potter_universal_2015} (Sec.~\ref{sec:transition}) also conclude for such broad distributions in the thermal phase, near the transition. However, as also pointed out in Ref.~\cite{luitz_ergodic_2017}, there are reasons to believe that the Griffiths regions may not be solely responsible for the observed sub-diffusions: similar varying power-laws have been observed in systems with quasiperiodic potentials~\cite{naldesi_detecting_2016,lev_transport_2017,lee_many-body_2017} --which do not harvest spatial Griffiths regions--,  and in two dimensional systems (in some approximate theoretical treatment~\cite{bar_lev_slow_2016} and in recent experiments~\cite{bordia_probing_2017}) where Griffiths effect should be less effective. Also at the quantitative level, the sub-diffusive effects are observed at quite low disorder (e.g. far off an estimate of the critical region), where one naively could expect Griffiths regions to have altogether minor influence. Other alternatives have been proposed, such as rare regions in the initial state~\cite{luitz_extended_2016} -- but this only applies for probes of slow dynamics after a quench.

As can be seen from this short summary, several questions regarding anomalous transport remain open. More work is needed in that direction, especially understanding the type of transport in dimension higher than one (which will perhaps soon no longer be out of reach for numerics). Another intriguing question is whether this sub-diffusive regime is related to the so-called 'bad metal' phase argued in Ref.~\cite{pino_nonergodic_2016}.

\subsection{Investigation in $d>1$}
\label{sec:dim}

So far, we mostly discussed MBL occurring in one-dimensional systems. Can a stable MBL phase be expected in dimension larger than one? This is a difficult question from the theoretical point of view, since exact numerics are not of much help in $d>1$ (due to the too large Hilbert spaces) whereas they are pivotal in $d=1$.  Furthermore, the RG treatments have not been adapted to dimension larger than one up to now. In this situation, we thus rely on more phenomenological scaling arguments.

It is long understood (see e.g. Ref.~\cite{aberg_onset_1990}) that the transition to the ETH phase can be roughly located by considering the ratio between the average matrix element and level spacing. Consider (as in e.g. Ref.~\cite{de_roeck_stability_2017,de_roeck_many-body_2017}) a thermal region in the sample of linear size $L$, surrounded by a localized region described in terms of lioms (Sec.~\ref{sec:liom}). The exponential many-body splitting scales as $\Delta  \sim \exp(-L^d).L^{d/2}$. The typical matrix element for flipping a liom located at a distance $r$ scales as $\Gamma  \sim \sqrt(\exp(L^d)). \exp(-r/\xi)$, where $\xi$ is some estimate of the localization length on the MBL side. When $\Gamma \gg \Delta$, the localized states hybridize (the liom flips) and the MBL region at distance $r$ from this ergodic grain is unstable and thermalizes.

This argument suggests that in dimension $d>1$,  localized systems are not stable. The competition between hybridization and many-body level spacing can be found in models of destabilization of MBL by ergodic grains or bubbles~\cite{de_roeck_stability_2017,de_roeck_many-body_2017,ponte_thermal_2017} and in phenomenological RG treatments~\cite{vosk_theory_2015,potter_universal_2015,dumitrescu_scaling_2017,thiery_many-body_2017}, and can be turned into  an effective criterion for locating the ergodic/MBL transition~\cite{serbyn_criterion_2015}. Of course, the argument above was refined in these works, by providing precise models for the thermal grain, the coupling with the MBL region, as well as backreaction effects on how the MBL regions can affect the spectral features of the ergodic grain (see Ref.~\cite{agarwal_rare-region_2017} for a discussion). Including all these ingredients appear to confirm that in dimension $d>1$, localized systems are unstable to inclusion of ergodic regions (caused e.g. by a rare fluctuation of disorder), even for very strong disorder. Indeed, up to now and despite numerical attempts (see e.g. the very interesting recent variational studies of Ref.~\cite{thomson_time_2018} and Ref.~\cite{wahl_signatures_2017}), there is no firm unbiased evidence of MBL in theoretical work in $d>1$ where the rigorous arguments of Ref.~\cite{imbrie_many-body_2016,imbrie_diagonalization_2016} do not apply, and the thermal phase is always present. 

Note finally that Ref.~\cite{chandran_many-body_2016} proposes, within an heuristic approach, that this description is incomplete and that in $d>1$, states can exist with at the same time ETH statistics and approximately conserved lioms, resulting in a dynamical phase similar to MBL. 

We emphasize that the argument above is asymptotic (valid for long-enough time and large systems), and that on time scales typical of those probed in e.g. cold-atom setups, experimental data in $d>1$ can be well explained/interpreted in terms of MBL and/or slow dynamics (see the 2d experiments presented in Sec.~\ref{sec:exp}). Finally, there might be models with some specific symmetries, disorder properties or geometrical constraints that could escape the argument (see Ref.~\cite{chen_how_2017} for an example of a 1d model with constraints that induce a MBL phase with a different mechanism). Also it does not apply for systems with long-range interactions (e.g. there is a MBL phase in the mean-field model of Ref.~\cite{laumann_many-body_2014,baldwin_many-body_2016}) -- see in this context the analysis of the localization transition in a central spin model~\cite{ponte_thermal_2017}.

\subsection{MBL as an Anderson problem on high-connectivity graphs} 
\label{sec:randomgraphs}

\subsubsection{Anderson problem in Fock and configuration spaces}
\paragraph{Fock space properties}
The picture of MBL as an Anderson problem on a complex graph has been developed early on in the first works on MBL~\cite{altshuler_quasiparticle_1997,gornyi_interacting_2005,basko_metalinsulator_2006}, relying on a weak interaction approach. In this formalism, a typical many-body Hamiltonian is rewritten as a tight-binding model defined on a complex graph of size ${\mathcal{D}}=$ dimension of the Hilbert space which grows exponentially with the volume\footnote{In this section and to make the discussion less specific to 1d, we use $N$ instead of $L$ for the total number of sites} $\sim \exp(N)$
\begin{equation}
H=-\sum_{\alpha}\mu_\alpha|\alpha\rangle\langle\alpha|-\sum_{\langle \alpha\neq \beta\rangle}V_{\alpha\beta}|\alpha\rangle\langle\beta| +{\rm{h.c.}}
\end{equation}
The states $|\alpha\rangle$ are many-body Slater determinants built on non-interacting localized orbitals, the on-site (correlated random) energies $\mu_\alpha$ are the Hartree-Fock terms, and the off-diagonal interaction part yields hopping $V_{\alpha\beta}$ between 'nearest-neighbour sites' $\langle \alpha\neq \beta\rangle$, see e.g. Ref.~\cite{bauer_area_2013} for more details. The number of matrix elements coupling two states $|\alpha\rangle$ and $|\beta\rangle$ grows polynomially with the volume $\sim N$ such that  it was argued by Bauer and Nayak that it is unlikely to observe a non-vanishing inverse participation ratio (IPR), defined for a given eigenstate $|e\rangle$ in the Fock basis by ${\rm{IPR}}_e=\sum_{\alpha}|\langle \alpha|e\rangle|^4$, even in the MBL regime. To our best knowledge, no such explicit numerical calculation of Anderson localization in the Fock space spanned by single particle Slater determinants has been reported. Note the related work of Refs.~\cite{bera_many-body_2015,bera_one-particle_2017} who consider the localization properties of the eigenstates of the single-particle density matrix computed in a given many-body eigenstate $|e\rangle$, giving insights on real space (but not Fock space) localization properties.

\paragraph{Configuration space}
Another way to explore the analogy between MBL and Anderson localization relies on rewriting the model as a single particle problem in a complete configuration basis, for instance for the $S=1/2$ Hamiltonian Eq.~(\ref{eq:rfH}) in the basis spanned by the spin projections $\{S_i^z\}$, where it reads 
\begin{equation}
H=-\sum_{k=1}^{\mathcal{D}} \mu_k|k\rangle\langle k|-t\sum_{\langle k l\rangle}|k\rangle\langle l|+{\rm{h.c.}},
\label{eq:heffAnderson}
\end{equation}
each 'site' $k$ encoding a given spin configuration $|\ldots\uparrow\downarrow\downarrow\uparrow\downarrow\ldots\rangle$.  For the particular many-body Hamiltonian Eq.~(\ref{eq:XXZ}), the on-site energies $\mu_k$ are given by the diagonal terms, i.e. for a given state $|k\rangle$ the local chemical potential is $\mu_k=\sum_{i=1}^{N}\langle k|h_iS_{i}^{z}-\Delta S_{i}^{z}S_{i+1}^{z}|k\rangle$, and the hopping between nearest neighbor sites $\langle k\,l\rangle$ is constant $t=1/2$ and given by the off-diagonal spin-flip terms of the original many-body Hamiltonian.  This analogy has been exploited in several works~\cite{monthus_many-body_2010,luca_ergodicity_2013,luitz_many-body_2015,pietracaprina_forward_2016,mondragon-shem_many-body_2015}. 

\begin{figure}[h!]
\begin{center}
\includegraphics[width=\columnwidth,clip]{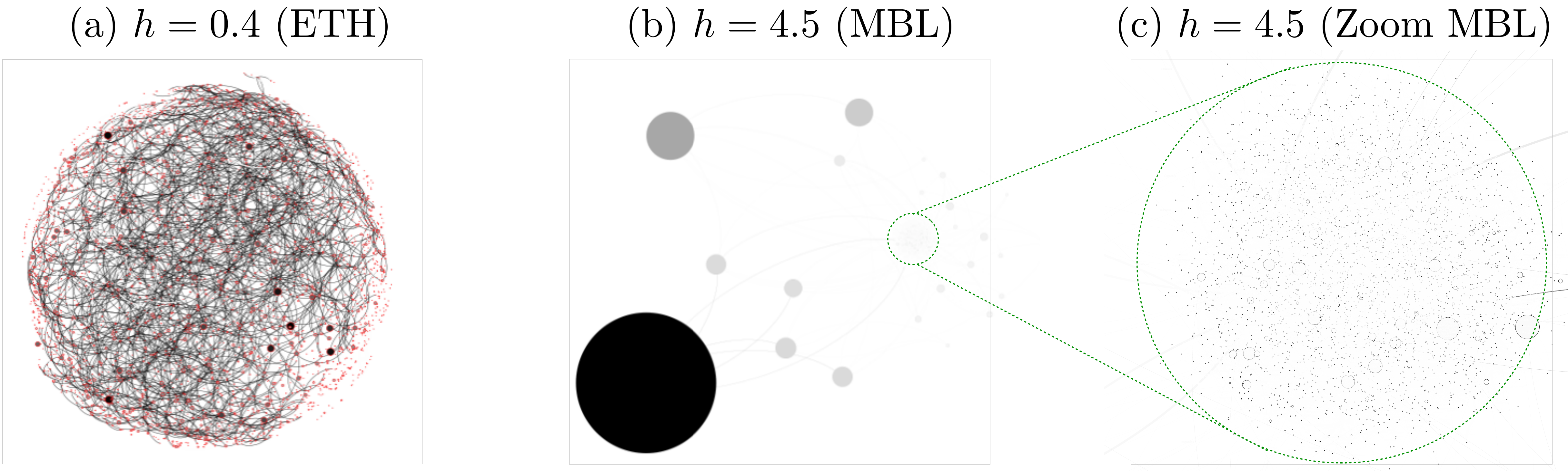}
\caption{Wave function occupation in the configuration space $\{|k\rangle\}$ of a single high-energy eigenstate $|\Psi\rangle$ (in the middle of the many-body spectrum) for the random-field Heisenberg chain model Eq.~(\ref{eq:rfH}). Results for chain of $L=16$ spins $S=1/2$, which corresponds to a $D=12870$ dimensional configuration space. In the ETH regime (a) one sees the weights $p_k=|\langle k|\Psi\rangle|^2$ almost uniformly distributed over the configuration space, with the links connecting nearest neighbors, the size of the circles being proportional to the weights $p_k$. In the MBL regime (b), we can distinguish one configuration which is strongly occupied (big black disk) together with a only few neighbors. Most of the configurations are occupied with a very small weight (region inside the green circle). Panel (c) is a zoom over this green circle region. While ergodicity is clearly visible in the ETH case (a), the non-ergodic character of the MBL eigenstate (b-c) is manifest.}\label{fig:participation}
\end{center}
\end{figure}

The explicit numerical calculation of the IPR in this computational basis was performed in Refs.~\cite{luca_ergodicity_2013,luitz_many-body_2015} where important insights have been reached regarding localization properties of eigenstates. A visual representation of typical eigenstates is given in Fig.~\ref{fig:participation}. In the delocalized regime (a) ergodicity is clearly visible with an almost perfect delocalization in the configuration space, i.e. ${\rm{IPR}}\sim 1/{\mathcal{D}}\sim 2^{-N}$. On the other hand, the MBL regime in panels (b-c) displays a clearly non-ergodic distribution of the weights, with a broad hierarchy between a 'localization center', its first neighbors, and further neighbors (see zoom in panel c)  which have very small weights. This breaking of ergodicity, first noticed by De Luca and Scardicchio in Ref.~\cite{luca_ergodicity_2013} has been confirmed for configuration space of size up to  ${\mathcal{D}}=705432$ by Luitz {\it{et al.}}~\cite{luitz_many-body_2015}. However, the support of MBL eigenstates is not finite, as one would naively expect for a simple Anderson localization in this complex network, but instead numerical data from~\cite{luitz_many-body_2015} rather suggest ${\rm{IPR}}\sim 1/N^l$ or ${\rm{IPR}}\sim \gamma^{-N}$ (with $\ln(\gamma) \ll 1$), probably a direct consequence of the connectivity of the configuration space which grows as $N$. \\

\paragraph{Strong disorder decimation in the configuration space} It is worth mentioning the original work of Monthus and Garel in Ref.~\cite{monthus_many-body_2010} where they generalized the Aoki real-space renormalization scheme for single particle Anderson localization~\cite{aoki_real-space_1980} to the above formulated many-body problem Eq.~(\ref{eq:heffAnderson}). By gradually decimating the sites having the strongest couplings, new couplings are generated in the configuration lattice up to the two last surviving sites. The finite size scaling of the effective hopping between these two last sites gives an estimate of the two-point Landauer transmission~\cite{monthus_statistics_2009,mard_strong-disorder_2017} which 
 decays exponentially with the distance in the MBL phase, while it remains finite in the delocalized regime. Using such an approximate scheme, a quite accurate estimate of the critical point was obtained~\cite{monthus_many-body_2010} for the interacting fermionic chains with nearest and next-nearest neighbour repulsion in a random potential, in good agreement with numerical results~\cite{oganesyan_localization_2007}.

\subsubsection{Links with Anderson localization and ergodicity breaking in random graphs}
The fact that MBL physics can be viewed as an Anderson problem defined on highly connected random graphs has triggered several debates addressing the possible connections with single particle dynamics on tree-like complex networks~\cite{altshuler_quasiparticle_1997,gornyi_interacting_2005,biroli_difference_2012,de_luca_anderson_2014,altshuler_nonergodic_2016,tarquini_critical_2017}. While a true localization transition is not questionnable on such finite connectivity graphs~\cite{abou-chacra_selfconsistent_1973,abou-chacra_self-consistent_1974}, the most pressing issue, still debated, concerns the possibility to stabilize a delocalized non-ergodic phase (dubbed 'bad metal'), intervening in a finite region of the phase diagram between insulating and conventional delocalized regime. Note that the simplest example of such a strange phase is the critical state at the Anderson transition point where non-ergodicity can be characterized by a set of multifractal exponents~\cite{evers_anderson_2008}. Various numerical works have reached different conclusions regarding the possibility to achieve such a non-ergodic bad metal in random graph geometries. First suggested in Refs.~\cite{de_luca_anderson_2014,altshuler_nonergodic_2016,altshuler_multifractal_2016}, it was later argued that this apparent lack of ergodicity was due to severe finite size effects~\cite{tikhonov_anderson_2016,tikhonov_fractality_2016,garcia-mata_scaling_2017}. In particular, using large system sizes (up to $2\times 10^6$ sites) Garcia-Mata {\it{et al.}}~\cite{garcia-mata_scaling_2017} identified a single delocalized regime which appears non-ergodic at small scale below a non-ergodicity volume (diverging at the localization transition), but ergodicity is recovered at large enough scale above this characteristic volume. 

\subsubsection{Consequences for the many-body problem}
While the structure of the configuration or Fock spaces locally ressembles a tree, there are several differences that could potentially change the physics: (i)  the connectivity grows slowly (logarithmically) with the size of the Hilbert space; (ii) the random energies are strongly correlated since an exponentially large number of $\mu_k$ in Eq.~(\ref{eq:heffAnderson}) are built from $N$ random fields ; (iii) the network is not precisely random but rather built deterministically by rules coming from the off-diagonal elements. As argued in Ref.~\cite{luca_ergodicity_2013}, the growing connectivity is probably the most serious ingredient which makes the approximation questionnable. 

Nevertheless, the question addressing the existence of a bad metal regime in the delocalized phase can also be posed for the many-body problem. Evidences for a unique delocalized phase (asymptotically ergodic in the configuration space) has been provided by Refs.~\cite{luitz_many-body_2015,serbyn_thouless_2017} while Refs.~\cite{pino_multifractal_2017,torres-herrera_extended_2017} conclude for an intermediate non-ergodic region. However, as discussed by Luitz and Bar Lev in Ref.~\cite{luitz_ergodic_2017} a precise definition of non-ergodicity is still lacking, and one should perhaps favor basis-independent measures of ergodicity, which is not the case with the IPR or related participation entropies. 

Finally, let us also mention the recent analysis on the Bethe lattice by Biroli and Tarzia~\cite{biroli_delocalized_2017} who report anomalous slow dynamics when approaching the Anderson localization transition, at least for the finite time range explored in their numerics. Interestingly, this observation supports the idea that studying the non-interacting Anderson model on the Bethe lattice as a toy model for MBL allows to qualitatively capture non-trivial features, such as the subdiffusion discussed in Section~\ref{sec:sub}.

\subsection{How to (experimentally) distinguish a MBL from an Anderson insulator? }
\label{sec:distinguish}

As presented in Sec.~\ref{sec:exp}, there is clear evidence reported for the existence of a dynamical localization in cold-atom experiments, but a vexing situation still persists. If we blindly consider the data of Fig.~\ref{fig:Imb} in the disordered regime without looking at the legend, there is no way to know if they emanate from an Anderson insulator or a MBL phase, as both phases result in a non-zero saturation value of imbalance. Showing that the localization observed in experiments is indubitably MBL would be very compelling. 

At the theoretical level, the difference between an Anderson insulator and a MBL insulator is subtle but well identified:  it emanates from the existence of interactions between the lioms (all terms behind the first one in Eq.~\ref{eq:Hliom}) in the MBL case. Physically, this results in dephasing being present in MBL, while it is not there for Anderson insulators. In this framework, the goal is in principle now simple: find a probe of dephasing. 

Several ways are possible. The first historical proof of dephasing was the slow growth of entanglement entropy after a quench in the MBL phase (for an Anderson insulator, the entanglement is rapidly bounded by a constant). While there have been important pioneer experiments~\cite{islam_measuring_2015,kaufman_quantum_2016}, the experimental determination of entanglement dynamics in a scalable many-body system is however not yet at a mature stage to be able to distinguish Anderson localization from MBL.

Still within the framework of quench experiments, one can find an experimentally-accessible observable $\langle O(t) \rangle$ that will contain a contribution from the time evolution of one liom operator $\langle \tau^x_i(t) \rangle $ (or a few, such as  $\langle \tau^x_i(t)\tau^x_j(t) \rangle $): this is actually generically the case for most local observables. Such observable will generically decay to its long-time value in a power-law fashion characteristic of the MBL phase (related to the lengthscale on which lioms interact), whereas it will not decay in an Anderson insulator~\cite{serbyn_quantum_2014}. There may be however several practical caveats. Oscillations on top of the power-law can be strongly depending on the local observable, the initial state and the final saturation value. Also the saturation value (which depends on the overlap of the local observable with the conserved quantity $\tau_i^z$) may be close to the initial value, and render the determination of a power-law regime difficult. These two effects are typically present in the imbalance of Sec.~\ref{sec:exp}. Other observables such as correlation fonctions at different sites may contain contributions from different power-laws, complicating the analysis~\cite{serbyn_quantum_2014}. In practice, a few observables have been demonstrated to qualify, most of them rooting in quantum information: the concurrence between nearest-neighbor particles~\cite{iemini_signatures_2016}, or the dynamics of the quantum mutual information between two~\cite{de_tomasi_quantum_2017} (or more~\cite{banuls_dynamics_2017}) sites.

Another possible route is interference experiments. A first work~\cite{serbyn_interferometric_2014} highlighted how a coherent spin manipulation at two different locations in the system (non-local spin-echo protocol) could pick up the weak interaction between the lioms, resulting in different signals for MBL and Anderson insulators. An alternative method based on spin noise spectroscopy has also been proposed~\cite{roy_probing_2015}. Finally, other schemes use a single-spin probe coupled to the disordered many-body system of interest, and allow to distinguish Anderson from MBL by considering revivals in dynamics of this probe~\cite{vasseur_quantumrevivals_2015}, or using the ability of this setup to measure the Loschmidt echo~\cite{serbyn_loschmidt_2017}.

\begin{figure}[t!]
\begin{center}
\includegraphics[width=0.6\columnwidth,clip]{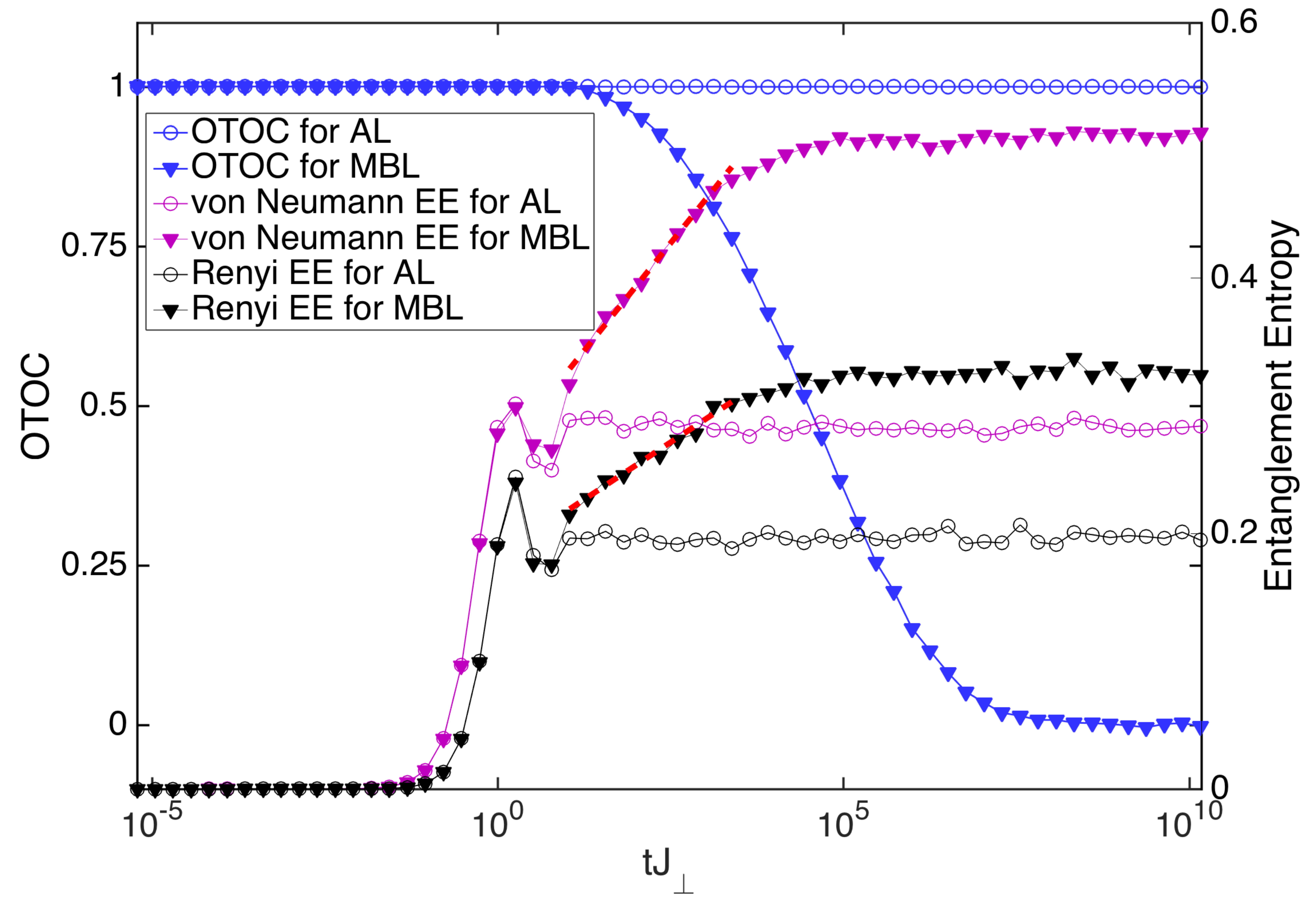}
\caption{From Fan {\it{et al.}}~\cite{fan_out--time-order_2017}: Dynamics after a quench from an initial product state of the entanglement entropy and of the OTOC (with $A=\sigma_2^x$, $B=\sigma_8^x$) for a random field XXZ chain Eq.~\ref{eq:XXZ} of size $L=8$., in the Anderson ($\Delta=0$) and MBL ($\Delta=0.2$) regimes for large disorder strength $h=5$. In the MBL phase, the OTOC decays as a power-law whereas it stays constant $\simeq 1$ in the Anderson case. Figure extracted from R. Fan {\it et al.}, Science Bulletin {\bf 62}, 707 (2017), under \href{https://creativecommons.org/licenses/by/4.0/}{Creative Commons Attribution License (CC BY)}.}
\label{fig:OTOC}
\end{center}
\end{figure}

Another type of correlator was also shown to be useful in this context. This is the Out-of-Time-Order Correlator (OTOC), defined as follows $C(t) = \langle A(t) B A(t) B \rangle$ where A and B are hermitian local operators (think about spin flips at different positions $A=S^x_i$ and $B=S^x_j$ for instance in the random field Heisenberg spin chain Eq.~\ref{eq:rfH}), and $A(t)=e^{iHt} A e^{-iHt}$. This correlator was introduced in the context of quantum chaos theory~\cite{shenker_black_2014,maldacena_bound_2016} as a measure of scrambling of quantum information, the impossibility of the system to recover information lost in dynamics solely by performing local measurements. Indeed, it can be understood as the overlap between two states (see e.g. Ref.~\cite{chen_out--time-order_2016}): one where we first punch the system by $B$, let the system evolve during time t to then punch with $A$, and a second state where we first let the system evolve to $t$ to apply $A$, then go back in time to apply $B$ at time zero, and finally let the system evolve back to time $t$ to be able to compare it with the first state. The OTOC is a quantum proxy of the 'butterfly effect' as it diagnoses the spread of quantum information, this feature turning out to be crucial to distinguish the various states of matter of interest. In an ergodic/chaotic system, the OTOC reaches its saturation value exponentially fast. On the other hand, there is no spread of information in an Anderson insulator and the OTOC never decays. Finally, in the MBL phase, the spread of information is slow, which will result in a power-law decay of OTOC to this saturation value. A series of paper showed clearly that the OTOC distinguishes MBL from Anderson~\cite{huang_out--time-ordered_2016,fan_out--time-order_2017,he_characterizing_2017,chen_universal_2016,swingle_measuring_2016,chen_out--time-order_2016}, as much as the dynamics of the EE do (see Fig.~\ref{fig:OTOC} for an illustration). 

While the definition of OTOC is more complex, it has probably a richer physical context. What about experiments? Several protocols have been proposed to measure the OTOC, where the needed reversal of time is often performed using an ancilla/switch/control degree of freedom~\cite{swingle_measuring_2016,yao_interferometric_2016,zhu_measurement_2016}.  As a matter of fact,  the OTOC has been experimentally measured for specific initial/thermal states in a molecule-based NMR quantum simulator~\cite{li_measuring_2017} and trapped ions~\cite{garttner_measuring_2017} systems. These protocols seem to be complex to be carried out in the currently working MBL experiments (Sec.~\ref{sec:exp}), and there is probably a need to find a closely related measure that could be experimentally implemented. A recent step in this direction has been proposed in Ref.~\cite{bordia_out_2018} where, instead of performing an overlap, a measurement is inserted after an OTOC sequence.
     
\subsection{The computational challenge of MBL}
\label{sec:challenge}

As can be seen above, numerical studies play a great role in the study of MBL. Conversely, the specificities of the MBL problem (physics in the middle of the spectrum, isolated systems, strong disorder, low entanglement) pushed computational physicists to create new methods or extend some existing ones. Indeed, well-established numerical methods for the condensed-matter many-body problem face an important challenge. As MBL physics primarily deals with isolated systems (no bath), usual finite-temperature quantum Monte methods, or high-temperature series expansion are not appropriate. Also, techniques which are designed to capture ground-state physics, such as iterative diagonalization (e.g. Lanczos method), standard density matrix renormalization group (DMRG) or zero-temperature series expansions, bring no information for the middle of the spectrum. 

Specific numerical methods have been used which work in principle in the full phase diagram of MBL models (see a partial review in Ref.~\cite{luitz_ergodic_2017}). Shift-invert exact diagonalization consists in applying Krylov space iterative methods to the operator $(H-\lambda)^{-1}$, allowing to probe an arbitrary energy density in the spectrum by computing the eigenpairs closest to the shift $\lambda$. These methods, used for the first time in the MBL context in Ref.~\cite{luitz_many-body_2015}, allowed to reach up systems composed of up to $22$ spins 1/2 for model Eq.~(\ref{eq:rfH}) (instead of $~16$ spins for standard full diagonalization methods). A recent extension~\cite{pietracaprina_shift_2018} pushed further the limit of shift-invert calculations to $26$ spins 1/2. The Krylov space is also very useful for exact time dynamics to expand the evolution operator $e^{-iHt}$, allowing to simulate exactly dynamics of up to $34$ spins 1/2 ~\cite{brenes_massively_2017,luitz_extended_2016}, albeit up to moderately small times (typically $t\sim 100$ in the units of Eq.~(\ref{eq:rfH}). This technique is thus particularly useful when thermalization is fast (at low disorder) and complementary to other methods described below which work well when the information spread is slow. Numerical linked cluster expansion techniques have also been extended to work for the MBL problem, using either discrete~\cite{tang_quantum_2015} or continuous~\cite{devakul_early_2015} disorder.

Another effective strategy is to implicitly or explicitely use properties/knowledge of the MBL phase to extend existing methods. The low entanglement of the MBL states allows them to be captured and computed as matrix product states~\cite{bauer_area_2013,khemani_obtaining_2015,yu_finding_2015}. In particular, the DMRG algorithm has been adapted to high-energy excited states by targeting the low-energy states of the shifted Hamiltonians  $(H-\lambda)^2$ or $(H-\lambda)^{-1}$ ~\cite{yu_finding_2015,kennes_entanglement_2016,lim_many-body_2016,serbyn_power-law_2016,villalonga_exploring_2018}. An interesting alternative (dubbed DMRG-X) is to take advantage of the liom structure and target a MPS state with maximal overlap with a simple product state~\cite{khemani_obtaining_2015,devakul_obtaining_2017}. Since in the full-MBL phase, all states have a matrix product state structure, it is possible to define the full diagonalization unitary operator as a Matrix Product Operator~\cite{pekker_encoding_2017,pollmann_efficient_2016,wahl_entire_2016,chandran_spectral_2015}, leading to numerical schemes which are efficient relatively deep in the MBL phase.  MPS methods are also very useful to follow the dynamics in the MBL phase due to the slow growth of entanglement: the already-existing time-evolving block decimation (TEBD), time-dependent DMRG (t-DMRG) or light-cone RG algorithms~\cite{andraschko_purification_2014} are particularly powerful in this context. 

The promising family of methods of unitary flows also takes explicitly advantage of the liom structure in the MBL phase. The idea is to iteratively produce transformations on a small scale that diagonalize the Hamiltonian -- in doing so, the lioms are explicitely constructed~\cite{rademaker_explicit_2016,rademaker_many-body_2017,thomson_time_2018}. This procedure can be seen as an example of a Hamiltonian flow towards its diagonalization, for which there are several possibilities~\cite{monthus_flow_2016}. A particular form, the Wigner flow, has been used to estimate the transition between MBL and ETH phases~\cite{pekker_fixed_2017}. These flow techniques have been extended to study time dynamics after a quench in the MBL phase~\cite{thomson_time_2018}. 

In a different interesting perspective, it was also proposed to construct local approximations to the lioms to define a shifted Hamiltonian which ground-state properties are studied using quantum Monte Carlo techniques~\cite{inglis_accessing_2016}. 

Finally, real-space renormalization group techniques usually considered for computing ground-state properties of strongly disordered systems were extended to finite-excited states (RSRG-X)~\cite{pekker_hilbert-glass_2014}, and have been instrumental to determine the properties of the phase transition inbetween two MBL phases (with or without order, see Sec.~\ref{sec:Ising}), of quantum critical glasses~\cite{vasseur_quantum_2015} as well as  to build the effective models~\cite{potter_universal_2015} for the RG approach to the transition (Sec.~\ref{sec:transition}). Whether or not one of these methods, a combination of several, or a completely new idea will be able to better capture the transition between the ETH and MBL phases is an open issue (see discussion in Sec.~\ref{sec:transition}). 

\section{MBL phase and quantum simulations}
\label{sec:qc}

As this is the main subject of this topical issue, we take this opportunity to briefly review connections between MBL and quantum simulations, in several possible meanings. The peculiar entanglement properties of MBL states (see Sec.~\ref{sec:intro},~\ref{sec:ent_transition} and~\ref{sec:ent}) naturally suggest such connections.

MBL states have been very early on argued to be good candidates to store quantum information~\cite{bauer_area_2013}. The main idea is that a MBL system has memory of its initial conditions and for instance the sign of an initially encoded imbalance will not change even at very long times, and will be measurably different from zero. Another possibility is to use MBL states which sustain topological/Ising order (see Sec.~\ref{sec:Ising}), which is actually protected by disorder from decoherence~\cite{huse_localization-protected_2013}.  Consequently, two quantum information processing protocols based on MBL states have been proposed: one explicitly uses the symmetry breaking mechanism of Sec.~\ref{sec:Ising} to store and manipulate (including with quantum gates) quantum information~\cite{yao_many-body_2015}, the other uses the MBL dephasing as a purification resource in a protocol based on spin echoes and quantum phase estimations~\cite{choi_quantum_2015}. 

 The operability of a quantum computing system based on MBL states would crucially depend on the smallness of the coupling to the environment. Bauer and Nayak~\cite{bauer_area_2013} argue that the MBL phase cannot act as a self-correcting quantum memory in the case of a non-zero external perturbation rate. Very early work~\cite{georgeot_quantum_2000} also focused on the destructive role of interactions (i.e. they induce a transition to the thermal phase) for a quantum-computing model based on a localized many-body system, even for a perfectly isolated system. Quite interestingly, taking an opposite point of view, Ref.~\cite{bauer_analyzing_2014,childs_toward_2017} analyze how a working  quantum computer (with a relatively small number of qubits) could help in simulating/understanding MBL on systems larger than those available to classical computations (Sec.~\ref{sec:challenge}), with Ref.~\cite{childs_toward_2017} discussing performances of various quantum algorithms to simulate Eq.~(\ref{eq:rfH}) on a quantum computer.
  
In this spirit, the cold-atom experiments presented in Sec.~\ref{sec:exp} can be seen as quantum analog simulations of theoretical models of MBL. At the moment, the standard model of MBL (the spin chain of Eq.~\ref{eq:rfH}) has never been implemented experimentally, but as already mentioned there are already several situations where the quantum analog experiments are further ahead than simulations (on classical computers), in particular for 2d systems or regarding controlled coupling to a bath. Of course, not all observables or properties of MBL can be measured given a specific quantum analog setup --the cold-atom experiments cannot compute properties of (large) individual  eigenstates, or spectral statistics--and this situation is certainly comparable to other quantum simulators. An interesting point that can be reached with the current experimental state-of-the-art would be to create specific disorder patterns (such as including Griffiths regions with, say,  a local strong disorder), in order to probe the bubble-like scenarios which are pivotal in the current understanding of MBL stability. Another relevant point would be to probe the existence of a mobility edge in experiments that probe 1d MBL, by e.g. preparing initial states with different energy densities, or by using different measurements that probe independently the existence of localized and extended states. The latter program was recently realized to demonstrate the existence of a single-particle mobility edge in a non-interacting system by measuring long-time imbalance and cloud expansion of a charge density-wave state prepared to be initially present in the center of an optical lattice~\cite{luschen_exploring_2017}. 
  
Finally, we would like to insist that the exploration of MBL physics triggered several important technical developments in numerical computations of interacting quantum systems, as reviewed in Sec.~\ref{sec:challenge}, and allowed to reveal new phases of matter (excited-state order, Floquet time crystals --see Sec.~\ref{sec:conc}--) which are not present in non-disordered systems. It thus sounds likely that in the near future new quantum computation schemes or implementations will be devised, both at theoretical and experimental levels, based on the unique properties of MBL.

\section{Conclusion}
\label{sec:conc}

We tried to highlight several aspects of the new physics appearing in closed quantum systems when interactions and strong disorder effects interplay, leading to new states of matter. The resulting many-body localized phase can be considered as the generic scenario beyond thermalization in disordered quantum systems. Quantum entanglement being a cornerstone to MBL, several interesting perspectives naturally emerge for alternative models of quantum computation with/for MBL states. 

There are several aspects of the problem not covered here. For instance, we did not discuss what happens when a MBL phase is coupled to a bath~\cite{nandkishore_spectral_2014,hyatt_many-body_2017,nandkishore_general_2017,znidaric_diffusive_2016}:  adding dissipation often leads to stretched exponential signals~\cite{levi_what_2015,fischer_dynamics_2016,carmele_stretched_2015,luschen_signatures_2016}, e.g. for imbalance decays. Another very interesting topic which we left out are the new 'Floquet' phases of matter, such as time crystals (see e.g.~\cite{else_floquet_2016,moessner_equilibration_2017,zhang_observation_2017,choi_observation_2016}), that emerge when a MBL system is periodically driven. Also a very interesting issue is the fate of MBL for systems with long-range interactions \cite{burin_energy_2006,hauke_many-body_2015, burin_many-body_2015,nandkishore_many_2017}, in particular dipolar interactions~\cite{yao_many-body_2014} relevant for recent experiments on e.g. NV centers in diamond~\cite{choi_observation_2016,kucsko_critical_2016}. 

We conclude by emphasizing that much more questions are still open or yet to be asked in this exciting topic.

\section*{Acknowledgements}
We gratefully acknowledge discussions with David Luitz and Gabriel Lemari\'e, and useful suggestions by Fran\c{c}ois Huveneers, C\'ecile Monthus, Antonello Scardicchio and Romain Vasseur. This work was supported by the French National Research Agency (ANR) under projects THERMOLOC ANR-16-CE30-0023-02, BOLODISS ANR-14-CE32-0018, and by Programme Investissements d’Avenir under the program ANR-11-IDEX-0002-02, reference ANR-10-LABX-0037-NEXT.

\bigskip

\bibliography{revue_final}
\bibliographystyle{elsarticle-num}


\end{document}